\documentclass[10pt,a4paper,english,oneside]{article}

\usepackage[margin=1.5in]{geometry}

\usepackage{authblk}

\usepackage[english]{babel}
\usepackage[utf8]{inputenc}   
\usepackage[T1]{fontenc} 
\usepackage{amsmath, amsfonts, amssymb, amsthm, mathrsfs, bm,bbm, mathtools,dsfont,yhmath}
\usepackage{braket}
\usepackage{enumerate}
\usepackage{mdwlist}
\usepackage{tensor}
\usepackage{empheq}
\usepackage[makeroom]{cancel}
\usepackage{simplewick} 

\usepackage{subcaption}

\usepackage[breaklinks]{hyperref}
\usepackage{xcolor}
\hypersetup{
    colorlinks,
    linkcolor={red!80!black},
    citecolor={green!70!black},
    urlcolor={blue!80!black}
}
\colorlet{dkgreen}{green!50!black}

\usepackage{tikz}

\usepackage[babel]{csquotes}
\usepackage[style=numeric-comp,backend=biber,mcite=true,doi=false,url=false,isbn=false,eprint=false,giveninits=true,sorting=none]{biblatex}
\bibliography{./bib}
\newbibmacro{string+doiurlisbn}[1]{%
  \iffieldundef{doi}{%
    \iffieldundef{url}{%
      \iffieldundef{isbn}{%
        \iffieldundef{issn}{%
          #1%
        }{%
          \href{http://books.google.com/books?vid=ISSN\thefield{issn}}{#1}%
        }%
      }{%
        \href{http://books.google.com/books?vid=ISBN\thefield{isbn}}{#1}%
      }%
    }{%
      \href{\thefield{url}}{#1}%
    }%
  }{%
    \href{http://dx.doi.org/\thefield{doi}}{#1}%
  }%
}
\DeclareFieldFormat{title}{\usebibmacro{string+doiurlisbn}{\mkbibemph{#1}}}
\DeclareFieldFormat[article,incollection,inproceedings]{title}%
    {\usebibmacro{string+doiurlisbn}{\mkbibquote{#1}}}

\renewcommand{\Pi}{\varPi}
\newcommand*{\diff}{\mathop{}\!\mathrm{d}}

\newcommand*\Diff[1]{\mathop{}\!\mathrm{d}^#1}

\DeclareMathOperator{\Tr}{Tr}

\DeclareMathOperator{\sgn}{sgn}

\DeclareMathOperator{\erfc}{erfc}

\newtheorem{CSP}{Constraint satisfaction problem}

\def\mean#1{\mathinner{\langle{#1}\rangle}}

\begin{document}
\title{Critical properties of the SAT/UNSAT transitions in the classification problem of structured data}

\author{Mauro Pastore}
\affil{\small\textit{Université Paris-Saclay, CNRS, LPTMS, 91405 Orsay, France} \\
mauro.pastore@u-psud.it
}

\maketitle
\begin{abstract}
The classification problem of structured data can be solved with different strategies: a supervised learning approach, starting from a labeled training set, and an unsupervised learning one, where only the structure of the patterns in the dataset is used to find a classification compatible with it. The two strategies can be interpreted as extreme cases of a semi-supervised approach to learn multi-view data, relevant for applications. In this paper I study the critical properties of the two storage problems associated with these tasks, in the case of the linear binary classification of doublets of points sharing the same label, within replica theory. While the first approach presents a SAT/UNSAT transition in a (marginally) stable replica-symmetric phase, in the second one the satisfiability line lies in a full replica-symmetry-broken phase. A similar behavior in the problem of learning with a margin is also pointed out.
\end{abstract}

\tableofcontents

\section{Introduction}
\label{sec:intro}

A supervised-learning classification task is an inference problem aiming at finding a function, selected from a certain hypothesis class (also called architecture) by a training algorithm, able to fit the labels of known examples (the patterns in the training set). Moreover, the target function is required to have a good generalization performance, that is the ability of predicting labels of patterns not seen during the learning protocol.

In this context, it is crucial to evaluate informative measures of the expressive power, or complexity, of the architecture, which is the number of different classifications that the functions in the hypothesis class can possibly realize. Indeed, classical results in Statistical Learning Theory bound the generalization performance of a training algorithm with some metrics of the complexity of the architecture: 
loosely speaking, the ``richer'' a model is, the more chances there are to find a good solution of the learning problem; on the contrary, if it is too rich, overfitting prevents good generalization. Some of these bounds can be proven to be independent on the probability distribution of the dataset to learn, depending on the expressive power of the architecture only~\cite{Vapnik:1999,Bousquet:2004,Luxburg:2011}.

Despite their elegance and generality, however, these data-independent results are usually of little interest in practice, as the bounds they provide are often quite loose in the regime relevant for applications~\cite{Cohn:1992,Zhang:2021}. As a possible way to overcome this drawback, in the last twenty years or so it was established, both in the mathematics~\cite{ShaweTaylor:1998} and in the statistical physics community~\cite{Sompolinsky:2018prx,Erba:2019,Zdeborova:2020prx,Abbaras:2020}, the distinguished role of data structure to understand the generalization performance of machine learning architectures.
The main idea behind this research line is that input-label correlation, modeling the fact that ``similar'' inputs (in the sense for example of invariance under a symmetry group) should have similar labels, reduces drastically the number of classifications that a certain algorithm can learn in practice, and so its effective complexity.

Keeping track of data correlation it is possible to combine, with the notion of ``realizable classification'' within a certain hypotesis class, the concept of ``admissible classification'', that is a classification which is coherent with the structure of data. In other words, an admissible classification is such that it assigns the same label to sufficiently similar, or correlated, inputs. In~\cite{GPR:2020PRL,GPR:2020PRE,Gherardi:2021}, it is conjectured that a measure of the number of the admissible classifications alone within a certain hypothesis class should be a better bound for the generalization error than the classic data-independent measures of complexity. Indeed, while the number of all the realizable classifications usually grows monotonically with the number of inputs to classify (more points, more ways to label them), the admissibility constraint adds a competing effect of excluded volume (more points, less space available for a separating surface enforcing an admissible classification), which is eventually dominant as the number of inputs grows.

Since the seminal works by E. Gardner~\cite{Gardner:1987,Gardner:1988a,Gardner:1988b}, we know a way to estimate the expressive power of a certain architecture with the tools of the statistical mechanics of disordered systems. Indeed, the learning problem can be recast in form of a constraint satisfaction problem (CSP), in which the parameters spanning the hypothesis class have to be chosen in order to comply with the classification of the inputs in the training set. In the setting where inputs and labels are chosen at random, one can test the complexity of the architecture fitting ``pure noise''. Indeed, as the number of patterns defining the constraints grows, this CSP exhibits a satisfiability (SAT/UNSAT) transition, which is the point where it stops having solutions, that coincides with the storage capacity of the hypothesis class (the maximum number of random uncorrelated inputs that functions in the class can correctly classify).

In this work, I reconsider this approach in a more general setting, formulated in order to keep track of a precise kind of data structure. The simple generative model of correlated inputs I will consider consists in multiplets of points at fixed distances sharing the same label; in this case, two different CSPs can be defined. The first, analogous to the one studied by Gardner, concerns the search for a solution in the parameter space of the hypothesis class 
realizing a \emph{typical admissible classification}, and was formulated in~\cite{Opper:1995}; I will call it, for the purpose of this paper, supervised CSP (sCSP). The other is obtained relaxing the request of typicality in the assignment of the labels: for a typical configuration of the inputs alone, is there an admissible solution of the classification problem that can be found adjusting \emph{both the parameters and the labels}? This last problem is meaningful only in the case of structured data, being always possible to find a realizable classification of isolated points if the labels can be freely assigned, and was formulated in~\cite{GPR:2020PRL,GPR:2020PRE}; I will call it unsupervised CSP (uCSP).

Indeed, while the first CSP has a natural interpretation in the setting of supervised learning, where a solution is required to fit the labels in the training set, the second one is in practice unsupervised, in the sense that the only feature to learn is the structure of the objects in input, not their specific labels; one could say that the two CSPs aim to solve the same classification problem with different strategies. However, one can think of exploiting the two approaches in parallel, in a semi-supervised setting, using the two pieces of information on the dataset, the fact that the objects have a label and the fact that they have a structure that the classification must comply with, to reduce the complexity of the problem.

This idea has been proven very powerful in the context of multi-view learning~\cite{Zhao:2017}, where each pattern comes in multiple ``views'' and can be thought as a multiplet of points (in general, in different spaces) with a common label: multilingual documents, multimedia files with a video stream and an audio track, images of 3D objects from different angles, to be classified according to their contents. In all these cases, single-view classifiers can be trained independently on each view of the dataset, but they must show consensus in labeling views of the same pattern. This compatibility condition, analogous to the admissibility concept I introduced above for the simpler geometrical data I will consider in this paper, can be used for example to reduce the ratio of labeled/unlabeled examples the classifiers need to achieve low error~\cite{Blum:1998}. I will say more on the interpretation of the present work in light of this connection in the following.

The aim of this paper is to study the critical properties of the two CSPs introduced above in the case of the linear binary classification problem of structured data (doublets of points sharing the same label), with the machinery provided by replica theory. The phase space of these problems is described by two parameters: the load $\alpha=p/n$, where $p$ is the number of doublets in $\mathbb{R}^n\times \mathbb{R}^n$ to classify (in the thermodynamic limit, $p$, $n \to \infty$, $\alpha$ fixed), and a parameter $\rho$ describing the correlation between the points inside the same doublet.

For the sCSP, I will show that the SAT/UNSAT line $\alpha_c(\rho)$ lies entirely in a replica-symmetric (RS) phase, meaning that the problem remains convex up to the transition; however, the RS ansatz is only \emph{marginally stable} for any value of $\rho$. From the physical point of view, marginal stability means that, approaching the transition from the UNSAT phase, where an Hamiltonian accounting for the unsatisfied constraints can be defined, minima have flat directions, so the spectrum of excitations is gapless~\cite{Franz:2015}. However, as the RS ansatz loses stability only at the transition, it is expected that a slight modification of the problem, for example adding a small positive margin to the constraints, is enough to push the SAT/UNSAT line in a strictly stable region.

On the other hand, for the uCSP replica symmetry breaking (RSB) occurs in the SAT phase for any value of $\rho$; by evaluating the RS breaking point and the slope of the RSB order parameter function $q(x)$ at the breaking point, I can conclude that the satisfiability transition line $\alpha_*(\rho)$ lies in a phase exhibiting full RSB. This means that the phase space of the solutions of the underlying problem is non-trivial: at the RSB transition, in the SAT phase below the satisfiability transition, it shatters into clusters and ergodicity is lost. Also this result presents marginal stability~\cite{MPV:1986}, in the sense I explained above; in this case, however, as the SAT/UNSAT line is deep inside the RSB region, this property is robust under infinitesimal perturbations of the problem.

These findings should be seen as complementary to the results from the mathematical theory of computational hardness. Indeed, while the sCSP is a convex problem, that even in the worst case is easy to solve, the uCSP is a non-linear non-convex problem, that is known to be NP-hard~\cite{Park:2017}. Replica theory, on the other hand, allows to state results for a typical instance of the constraints, and the continuous nature of the RSB transition can be related to a precise scaling behaviour of the computing time needed to solve tipically the problem~\cite{Martin:2001,Monasson:1999}.

This paper is organized in the following way: in Sec.~\ref{sec:volumes} the two CSPs and their associated volumes of solutions are introduced; in Sec.~\ref{sec:multiFullRSB} I present the relevant equations from a full RSB replica approach in order to study the two problems; in Sec.~\ref{sec:Stability} I perform the stability analysis of the RS ansatz in the two problems, when their satisfiability transition is approached; in Sec.~\ref{sec:margin} some level of universality of these results, exploiting the analogy with the problem of learning with a margin, is highlighted; in Sec.~\ref{sec:semisup}, I explain, as a proof of concept, how to exploit data structure and the admissibility constraint in order to reduce the storage capacity of the (data dependent) hypothesis class in a semi-supervised approach, possibly motivating the current analysis in a way interesting for applications. Sections~\ref{sec:Stability}, ~\ref{sec:margin} and ~\ref{sec:semisup} collect the main results of this paper: first of all, marginal stability is found at the satisfiability transition of the CSPs related to the classification problem of structured data; moreover, the critical capacity of a linear classifier required to be compatible both to labeled and to unlabeled data, in a semi-supervised setting, is lower than the one of a pure supervised classifier. Interpretations and outlooks of the present work are reported in Sec.~\ref{sec:end}.

\section{Synaptic volumes for linear classification of structured data}
\label{sec:volumes}

In this section I introduce for reference the CSPs at the basis of the learning problem of correlated patterns, as explained in~\cite{GPR:2020PRE}.

The patterns to classify are $p$ doublets of points $\bm{\xi}=(\xi,\bar{\xi})$, each of them chosen uniformly on the vertices of an hypercube $\{-1,1\}^n$, but with fixed overlap $\rho$:
\begin{equation}
\frac{1}{n}\sum_{j}(\xi_j)^2 = \frac{1}{n}\sum_{j}(\bar{\xi}_j)^2 = 1\,,\qquad \frac{1}{n}\sum_{j}\xi_j \bar{\xi}_j = \rho\,.
\label{eq:doublets}
\end{equation}
Notice that most of the results presented here are valid also for multiplets of $k$ points correlated with a fixed correlation matrix $\mathcal{R}$, but I restrict to doublets for simplicity and to perform the calculation; the analytical formulas in this and the next sections can be generalized trivially. Points in the same doublet are required to share the same label $\sigma^\mu =\pm 1$: this condition defines the admissible dichotomies (binary classifications); apart from that, the labels are chosen at random uniformly in $\{-1,1\}$.

An architecture describing a linear classifier is parametrized by a weight vector $W\in\mathbb{R}^n$, assigning to each point $x\in\mathbb{R}^n$ a label $\sigma$ according to the rule $\sigma = \sgn (\sum_{j=1}^n W_jx_j)$. In the following, I will chose $W$ on the sphere $S^n(\sqrt{n})$. Following the seminal work~\cite{Gardner:1988a}, the problem of learning the values of $W$ in order to fit the training set can be cast as a constraint satisfaction problem, counting the typical number of planes realizing the linear classification. Taking into account the correlation between points, and thus the admissibility constraint, at least two different CSPs can be formulated in the present setting:
\begin{CSP}[supervised CSP, sCSP]
\label{problem:LSO}
Given a set of input-label pairs $\{\bm{\xi}^\mu, \sigma^\mu\}$
(with $\mu=1,\cdots,p$), find
a vector $W$ such that
\[
\sgn\!\left(\sum_{j=1}^n W_j \xi^\mu_j\right) = \sgn\!\left(\sum_{j=1}^n W_j \bar{\xi}_j^\mu \right) = \sigma^\mu \qquad \forall\,\mu\,.
\] 
\end{CSP}
\noindent In this problem, both the patterns and the labels take the role of quenched disordered variables; the statistical physics perspective can address questions about the volume of solutions for a typical instance of the set $\{\bm{\xi}^\mu, \sigma^\mu\}$.
\begin{CSP}[unsupervised CSP, uCSP]
\label{problem:GPR}
Given a set of inputs $\{\bm{\xi}^\mu\}$,
(with $\mu=1,\cdots,p$), find
a set of labels $\{\sigma^\mu\}$ and
a vector $W$ such that
\[
\sgn\!\left(\sum_{j=1}^n W_j \xi_j^\mu \right) = \sgn\!\left( \sum_{j=1}^n W_j \bar{\xi}_j^\mu \right)=\sigma^\mu \qquad  \forall \,\mu\,.
\]
\end{CSP}
\noindent Here, as explained in~\cite{GPR:2020PRL,GPR:2020PRE}, the labels are moved from the set of parameters defining the problem to the set of unknowns, and only the patterns $\{\bm{\xi}^\mu\}$ take the role of an instance of the quenched disorder. Note that this problem is not formulated in a supervised learning setting: as the labels can be freely adjusted to find a solution, the only thing that matters is the data structure enforcing the admissibility constraint on the dichotomy: two points in a doublet must be classified in the same way, no matter which one.

%
%
%

\begin{figure}
\centering
\includegraphics[width=.4\textwidth]{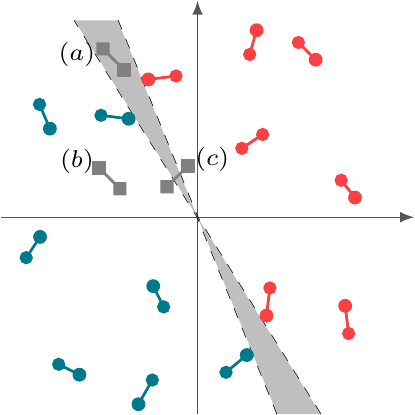}
\caption{A representation of the combinatorial reasoning used in~\cite{GR:2020PRR} to count the number of admissible linear dichotomies of doublets. The correlated inputs in Eq.~\eqref{eq:doublets} are drawn (pictorially) as extrema of segments of length $\sqrt{2n(1-\rho)}$. The admissible dichotomies, assigning a color (pink/blue) to each point, are the ones that do not break any segment; the dichotomy of $p$ doublets represented in figure is realized by any plane spanning the gray region. When a $(p+1)$th doublet (gray squares) is added, it can end up in one of the three kind of places labeled in figure. In $(a)$, it can be classified in two ways, choosing slightly different planes in the grey region; in $(b)$, it can be classified only in one way for any choice of plane in the gray region; in $(c)$ it invalidates the dichotomy. Along these lines, a mean-field recurrence for the number of admissible linear dichotomies $C_{n,p}$ can be written.}
\label{fig:combinatorics}
\end{figure}

In the following I will introduce the volumes counting the number of solutions of these two CSPs; however, let me spend some words on why they are interesting to study. First of all, the number $C_{n,p}$ of linear admissible dichotomies of $p$ doublets in $n$ dimensions is not trivial: the total number of dichotomies is $2^{2p}$, the admissibility constraint reduces it to $2^p$ (once a label for a point $\xi^\mu$ is chosen, the one for its partner $\bar{\xi}^\mu$ is fixed), but in general not all of them are linear (realized by hyperplanes separating the points with different labels). Indeed, when the doublets are sufficiently sparse ($p\ll n$), together with the fact that they are in random position, one can think that for any assignment of labels it is always possible to find a plane separating them, because a lot of directions are unconstrained, and so $C_{n,p}/2^p \approx 1$; but when $p$ is of the same order of $n$, this reasoning does not apply: if a doublet $\bm{\xi}^\mu$ carries a positive label $\sigma^\mu=1$, then any doublet close to $-\bm{\xi}^\mu$ must have a negative label in order for the classification to be linearly realizable.

In the case of the structured data defined above, it is possible to find $C_{n,p}$ via a combinatorial approach extending the classical result~\cite{Cover:1965}, which is valid only for isolated points in general positions, introduced in~\cite{GR:2020PRR} and summarized pictorially in Fig.~\ref{fig:combinatorics}. This combinatorial approach is dual to the two CSPs introduced above, which capture different aspects of the asymptotic behavior of $C_{n,\alpha n}$ for large $n$: the value of $\alpha$ where this number becomes subexponential in $n$ ($C_{n,\alpha n}/2^{\alpha n} \to 0$ for $\alpha>\alpha_c$) coincides with the critical SAT/UNSAT value of the sCSP, because above that a randomly chosen dichotomy will be typically not linearly separable; on the other hand, the value of $\alpha$ at which this number passes from growing to decreasing asymptotically ($C_{n,\alpha n} \to 0$ for $\alpha>\alpha_*$), due to the fact that the doublets are becoming too dense to leave enough space for a plane not cutting any of them, coincides to the critical SAT/UNSAT value of the uCSP, because then no possible linear admissible dichotomy exists.

According to the definitions above, the synaptic volumes for these CSPs, accounting for their number of solutions, are:
\begin{align}
V_{\text{s}} 
&=\! \int\! \left[ \prod_{j=1}^n \diff W_j \right]\!\delta\! \left(\sum_{j=1}^n W_j^2 -n \right)
\!\prod_{\mu=1}^{p}  \theta\!\left(\frac{1}{\sqrt{n}} \sum_{j=1}^n W_j \xi_{j}^\mu  \right)\theta\!\left(\frac{1}{\sqrt{n}} \sum_{j=1}^n W_j \bar{\xi}_{j}^\mu  \right)\,,\label{eq:volume_LSO}\\
V_{\text{u}} 
&= \!\int\! \left[ \prod_{j=1}^n \diff W_j \right]\!\delta\! \left(\sum_{j=1}^n W_j^2 -n \right)
\!\prod_{\mu=1}^{p} \sum_{\sigma^\mu = \pm 1}\!\! \theta\!\left(\frac{\sigma^\mu}{\sqrt{n}} \sum_{j=1}^n W_j \xi_{j}^\mu  \right)\theta\!\left(\frac{\sigma^\mu}{\sqrt{n}} \sum_{j=1}^n W_j \bar{\xi}_{j}^\mu  \right) \,. \label{eq:volume_GPR}
\end{align}
The volume~\eqref{eq:volume_LSO} is the analogous of the Gardner volume for this dataset, counting the number of planes, described by their normal vectors $W$, realizing a typical \emph{admissible} dichotomy: the input-output correlation is implemented requiring that the points in a doublet share the same label (without any loss of generality, $\sigma^\mu = +1$), and only those classifications complying with this condition are counted; I will say more on the analogy with Gardner's result in Sec.~\ref{sec:margin}.
The second volume is associated to the unsupervised satisfiability problem, where not only the planes but also the labels of the doublets can be adjusted to find a dichotomy.
Equivalently, for patterns forming doublets I can write
\begin{equation}
\begin{aligned}
V_{\text{u}} 
&= \int \left[ \prod_{j=1}^n \diff W_j \right]\delta\! \left(\sum_{j=1}^n W_j^2 -n \right)
\prod_{\mu=1}^{p}  \theta\!\left(\frac{1}{n} \sum_{i,j=1}^n W_i W_j \xi_{i}^\mu  \bar{\xi}_{j}^\mu  \right),\\
\end{aligned}
\end{equation}
where I used the step function identity $\theta(x)\theta(y) + \theta(-x)\theta(-y) = \theta(xy)$. The meaning of this simple observation is that the uCSP is a \emph{non-linear} (quadratically constrained) problem, in the sense that the vector of solution $W$ enters in a quadratic way in the unequality constraints. Similar problems (but with a different distribution of the disorder) have been recently studied from the statistical physics perspective in~\cite{Yoshino:2018}.

With standard manipulations~\cite{Duplantier:1981} in replica theory, valid in the thermodynamic limit $n$, $p\to \infty$, $\alpha = p/n$ fixed, I can write both the replicated-averaged volumes defined above in the compact form (see App.~\ref{app:replicas} for short reference)
\begin{multline}
\mean{V ^t}
= \int \left[\prod_{a<b} \diff Q_{ab} \right] \exp\Biggl\{\frac{nt[1+\log (2\pi)]}{2} + \frac{n}{2} \log \det Q \\
+ n\alpha \log \!\left[ \left.e^{\frac{1}{2}\sum_{a,b} Q_{ab}D_{ab}} \prod_a v(\bm{h}_a )\right|_{\{\bm{h}_a\}=0}\right]\Biggr\}\,,
\label{eq:volume_diff}
\end{multline}
where the brackets represent the average over the input distribution in the ensemble~\eqref{eq:doublets}, boldface letters indicate bidimensional vectors, $Q_{ab} = \sum_j W_j^a W_j^b/n$ is the $t\times t$ positive semi-definite overlap matrix between two replicas $a$ and $b$, and the functions $v(\bm{h})$ are, respectively,
\begin{equation}
v_{\text{s}}(\bm{h}) = \theta(h)\theta(\bar{h})\,,\qquad\qquad
v_{\text{u}}(\bm{h}) = \theta(h\bar{h})\,.
\end{equation}
Moreover, I introduced the following notation for the differential operator inside the logarithm:
\begin{equation}
D_{ab} =  \left(\frac{\partial}{\partial h_a}\,,\, \frac{\partial}{\partial \bar{h}_a} \right) \begin{pmatrix}
1&\rho\\
\rho&1
\end{pmatrix} 
\left(
\frac{\partial}{\partial h_b}\,,\, \frac{\partial}{\partial \bar{h}_b}
\right)^T\, = \bm{\partial}_a^T \mathcal{R} \bm{\partial}_b\,.
\label{eq:defR}
\end{equation}
I will call the vector $\bm{h}$ entering in the constraints ``gap variable'', following the literature on the subject. The intensive action from Eq.~\eqref{eq:volume_diff} is defined as
\begin{equation}
S(Q) = \frac{1}{2} \log \det Q 
+ \alpha \log \!\left[ \left.e^{\frac{1}{2}\sum_{a,b} Q_{ab} D_{ab}} \prod_a v(\bm{h}_a)\right|_{\{\bm{h}_a\}=0}\right]\,.
\label{eq:S}
\end{equation}
In the thermodynamic limit this quantity has to be evaluated at the saddle point for the elements of the matrix $Q$, defined by the equation:
\begin{equation}
0=\frac{\partial S}{\partial Q_{ab}} = \frac{1}{2}\left(Q^{-1}\right)_{ab}
+ \frac{\alpha}{2} \frac{\left. D_{ab} e^{\frac{1}{2}\sum_{a,b} Q_{ab} D_{ab}} \prod_a v(\bm{h}_a)\right|_{\{\bm{h}_a\}=0}}{\left.e^{\frac{1}{2}\sum_{a,b} Q_{ab} D_{ab}} \prod_a v(\bm{h}_a)\right|_{\{\bm{h}_a\}=0}}\,.
\label{eq:Qsaddle}
\end{equation}
Moreover, once an ansatz on the form of the replica matrix $Q$ is chosen, the eigenvalues of the Hessian matrix $\partial^2 S/\partial Q_{cd} \partial Q_{ab}$
have to be studied to check the stability of the solution at the saddle point. In the next section, I prefer to follow~\cite{Franz:2017scipost} working directly in the full RSB approach, and postponing the question of the stability of the RS ansatz to Sec.~\ref{sec:Stability}. Indeed, once the full RSB machinery is at hand, this analysis can be performed almost for free exploiting the marginally stable nature of the full RSB solution.

\section{Multidimensional form of the full RSB ansatz}
\label{sec:multiFullRSB}

In this section I derive the relevant expressions to study the solution of the saddle point equation~\eqref{eq:Qsaddle} and its stability in the full RSB scheme. In this approach, a hierarchical ansatz of the replica matrix $Q$ is taken (see App.~\ref{app:hierarchicalRSB} for details), breaking in steps the symmetry under permutation of any two replica indices; then, a certain limit for which these steps become infinitely many is performed. Accordingly, instead of obtaining a discrete set of values parametrizing the matrix $Q$, the solution is described by a continuous function $q(x)$ (the so-called order parameter function), with $x\in[0,1]$. 

While the general recipe to perform this calculation is known since~\cite{Parisi:1980} and can be found many times in literature, I report here the main results in the case studied in this paper, which present non-trivial features due to the multidimensional nature of the gap variables appearing in Eq.~\eqref{eq:volume_diff}. Even though I will not solve explicitly the problem in the full RSB scheme, I will report all the relevant equations I will need for the stability analysis of the RS ansatz, performed in the following section.

The free energy in the full RSB scheme for the present problem is the functional (see App.~\ref{app:FRSB} for an explicit derivation)
\begin{equation}
\begin{aligned}
S[x(q)]={}&{} \frac{1}{2}\left[ \log(1-q_M)+\frac{q_m}{\lambda(q_m)}+\int_{q_m}^{q_M} \frac{\diff q }{\lambda(q)} \right] +\alpha \left.\gamma_{q_m} \!\star\! f(q_m,\bm{h})\right|_{\bm{h}=0}\\
{}&{}-\alpha \int \Diff2 \bm{h} \,P(q_M,\bm{h}) \left[f(q_M,\bm{h}) - \log \gamma_{1-q_M}\!\star\! v(\bm{h}) \right]\\
{}&{}+\alpha \int \Diff2 \bm{h} \int_{q_m}^{q_M} \diff q \,P(q,\bm{h}) \biggl(\dot{f}(q,\bm{h})\\
{}&{}\hspace{6em} +\frac{1}{2} \left\{ \bm{\partial}^T\mathcal{R}\bm{\partial} f(q,\bm{h})  + x(q)\left[\bm{\partial}f(q,\bm{h})\right]^T\mathcal{R}\left[\bm{\partial}f(q,\bm{h})\right] \right\} \biggr)
\,,
\end{aligned}
\label{eq:FRSB_functional}
\end{equation}
where $x(q)$ is the inverse of the order parameter function $q(x)$, $q_m$ and $q_M$ are the minimum and maximum values of $q(x)$, the functional $\lambda$ is
\begin{equation}
\lambda(q)  =  1- q_M+ \int_q^{q_M} x(p) \diff p\,,
\label{eq:FRSB_lambda}
\end{equation}
and the Gaussian convolution is defined by
\begin{equation}
\gamma_{q} \!\star\! g(\bm{h}) = \int \frac{\Diff2 \bm{k}}{
2\pi q \sqrt{\det \mathcal{R}}} e^{-\frac{1}{2 q} \bm{k}^T \mathcal{R}^{-1} \bm{k}} g(\bm{h} + \bm{k})
= e^{\frac{q}{2} \bm{\partial}^T \mathcal{R} \bm{\partial}} g(\bm{h})\,.
\label{eq:convolution}
\end{equation}
The first line of Eq.~\eqref{eq:FRSB_functional} is the translation of Eq.~\eqref{eq:S} in the full RSB approach. The functional $f$, coming from the continuous limit of the hierarchical RSB ansatz, must have for consistency a precise structure implemented in $S$ via the functional Lagrange multiplier $P(q,\bm{h})$, following a variational approach devised in~\cite{Sommers:1984}.
Indeed, variations with respect to $P(q_M,\bm{h})$ and $P(q,\bm{h})$ give the \emph{Parisi equation} for $f$:
\begin{equation}
\left\{
\begin{aligned}
&f(q_M,\bm{h}) = \log \gamma_{1-q_M}\!\star\! v(\bm{h})\,,\\
&\dot{f}(q,\bm{h}) = -\frac{1}{2} \bigl\{\bm{\partial}^T\mathcal{R}\bm{\partial} f(q,\bm{h})  + x(q)\left[\bm{\partial}f(q,\bm{h})\right]^T\mathcal{R}\left[\bm{\partial}f(q,\bm{h})\right] \bigr\}\,,\quad q_m\le q \le q_M\,.
\end{aligned}\right.
\label{eq:FRSB_Parisi}
\end{equation}
Variations with respect to $f(q_m,\bm{h})$ and $f(q,\bm{h})$ (integrating by parts) give the equation of motion for the Lagrange multiplier:
\begin{equation}
\left\{\begin{aligned}
&P(q_m,\bm{h}) = \gamma_{q_m}\!(\bm{h}) = \frac{e^{-\frac{1}{2q_{m}}\bm{h}^T \mathcal{R}^{-1} \bm{h}}}{2\pi q_m \sqrt{\det{\mathcal{R}}}}\,,\\
&\dot{P}(q,\bm{h}) = \frac{1}{2} \left\{\bm{\partial}^T\mathcal{R}\bm{\partial} P(q,\bm{h}) - 2 x(q) \bm{\partial}^T \mathcal{R} \left[P(q,\bm{h}) \bm{\partial} f(q,\bm{h}) \right]  \right\}\,,\quad q_m \le q \le q_M\,.
\end{aligned}\right.
\label{eq:FRSB_multipliers}
\end{equation}
Variation with respect to the order parameter function $x(q)$ gives 
\begin{equation}
\frac{q_m}{\lambda(q_m)^2} + \int_{q_m}^q \,\frac{ \diff p}{\lambda(p)^2} = \alpha \int \Diff2 \bm{h}\, P(q,\bm{h}) \left[\bm{\partial}f(q,\bm{h})\right]^T\mathcal{R}\left[\bm{\partial}f(q,\bm{h})\right]\,.
\label{eq:FRSB_saddle1}
\end{equation}
Useful informations can be obtained differentiating this equation with respect to $q$. Indeed, with a first derivative I get an equation for $\lambda$
\begin{equation}
\frac{1}{\lambda(q)^2} = \alpha \int \Diff2 \bm{h} \,P(q,\bm{h}) \Tr\{[\mathcal{H}(q,\bm{h})\mathcal{R}]^2\}\,,
\label{eq:FRSB_saddle2}
\end{equation}
where $\mathcal{H}$ is the Hessian matrix
\begin{equation}
\mathcal{H}(q,\bm{h}) = \begin{pmatrix}
\partial_{hh}	& \partial_{h\bar{h}}\\
\partial_{h\bar{h}} & \partial_{\bar{h}\bar{h}}
\end{pmatrix}f(q,\bm{h})
\end{equation}
and $\Tr$ is the trace over the two-valued indices labeling the points in a multiplet (in the following, I will use Greek letters to represent them). Deriving another time I obtain an equation for the order parameter function $x(q)$, namely
\begin{equation}
x(q) = \frac{\lambda(q)}{2} \frac{\displaystyle\int \Diff2 \bm{h} \,P(q,\bm{h}) [\partial_\alpha \partial_\beta\partial_\gamma f(q,\bm{h})]\mathcal{R}_{\alpha\alpha'}\mathcal{R}_{\beta\beta'} \mathcal{R}_{\gamma\gamma'}[\partial_{\alpha'} \partial_{\beta'} \partial_{\gamma'} f(q,\bm{h})]}{
\displaystyle\int \Diff2 \bm{h} \,P(q,\bm{h})\left(
\Tr\{[\mathcal{H}(q,\bm{h})\mathcal{R}]^2\} + \lambda(q) \Tr\{[\mathcal{H}(q,\bm{h})\mathcal{R}]^3\}\right)}\,,
\label{eq:FRSB_saddle3}
\end{equation}
where the Greek indices are contracted. Deriving again I get an equation for the slope of the function $x(q)$:
\begin{equation}
\dot{x}(q)
\\
 = \frac{2x(q)}{\alpha B(q)}\left[\frac{\alpha}{2} \int \Diff2 \bm{h}\,P(q,\bm{h}) A(q,\bm{h}) - \frac{3 x(q)^2}{\lambda(q)^4}\right]\,.
\label{eq:FRSB_saddle4}
\end{equation}
with
\begin{align}
A(q,\bm{h}) &= [\partial^4_{\alpha\beta\gamma\delta} f(q,\bm{h} )]\mathcal{R}_{\alpha\alpha'}\mathcal{R}_{\beta\beta'} \mathcal{R}_{\gamma\gamma'}\mathcal{R}_{\delta\delta'}[\partial^4_{\alpha'\beta'\gamma'\delta'}f(q,\bm{h})] \notag\\
&\quad- 12 x(q) [\partial^3_{\alpha \beta \gamma } f(q,\bm{h})]\mathcal{R}_{\beta \beta'} \mathcal{R}_{\gamma\gamma'}[\partial^3_{\beta' \gamma'\alpha'} f(q,\bm{h})]\mathcal{R}_{\alpha'\delta}[\partial^2_{\delta \delta'} f(q,\bm{h})]\mathcal{R}_{\delta'\alpha} \notag\\
&\quad+6 x(q)^2 \Tr \{[\mathcal{H}(q,\bm{h})\mathcal{R}]^4\}\\
B(q) &=  \int \Diff2 \bm{h} P(q,\bm{h}) [\partial^3_{\alpha\beta\gamma} f(q,\bm{h}) ]\mathcal{R}_{\alpha\alpha'}\mathcal{R}_{\beta\beta'} \mathcal{R}_{\gamma\gamma'}[\partial^3_{\alpha'\beta'\gamma'}f(q,\bm{h})]
\end{align}
Eqs.~\eqref{eq:FRSB_functional}-\eqref{eq:FRSB_saddle4} are multidimensional generalizations of the stationary equations arising in the full RSB scheme, which are PDEs in a ``time'' variable $q$ and a single ``space'' variable $h$. Here, the covariance matrix $\mathcal{R}$ introduced in Eq.~\eqref{eq:defR} has the role of a metric tensor for the spatial part. $(k+1)$-dimensional formulas have already been reported in literature: in~\cite{Franz:2019multi}, $k$ is the number of hidden units in a two-layers, feed-forward neural network, and the metric tensor is the identity matrix; in~\cite{Gyorgyi:2001} (App.~E), general formulas arising from replica matrices with additional, internal indices are presented, equivalent to the ones I derived here.

I am now in the position to study the stability of the RS ansatz in the phase space of the two problems introduced in Sec.~\ref{sec:volumes}.

\section{The RS solution and its stability}
\label{sec:Stability}

The stability analysis of the RS ansatz is performed via the following reasoning~\cite{Franz:2017scipost}. In the hypothesis that the model presents full RSB, one can tune the control parameters ($\alpha$, $\rho$) to sit in a replica symmetry broken region arbitrarily close to the transition. The most natural way to break replica symmetry continuously is assuming that the order parameter function $q(x)$ at the transition is becoming non-constant around a point $x_m \approx x_M$, where $q_m\approx q_M$. This means that one can expand the formulas in the previous section at first order in $\dot{q}(x)$ at $x_m$. In particular, as we will see in the following, Eq.~\eqref{eq:FRSB_saddle2}, evaluated on the solutions of Eq.~\eqref{eq:FRSB_saddle1}, gives the condition of marginal stability of the RS ansatz at the transition. Moreover, Eqs.~\eqref{eq:FRSB_saddle3} and~\eqref{eq:FRSB_saddle4} can be used to find the values of $x_m$, the breaking point of the RS ansatz, and $\dot{q}(x_m)$, the slope of the order parameter function at the breaking point. In case these values are not consistent with the starting hypothesis of continuous RSB (for example, $x_m \notin [0,1]$, or $\dot{q}(x_m)<0$), the transition can be traced back to a discontinuous, stepwise RSB.

\begin{figure}[t]
\centering
\includegraphics[scale=1]{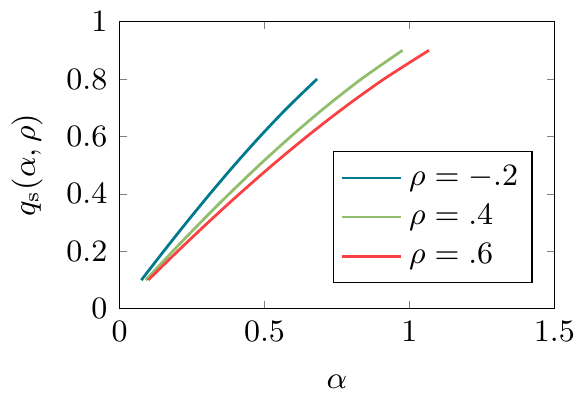}
\hspace{2em}
\includegraphics[scale=1]{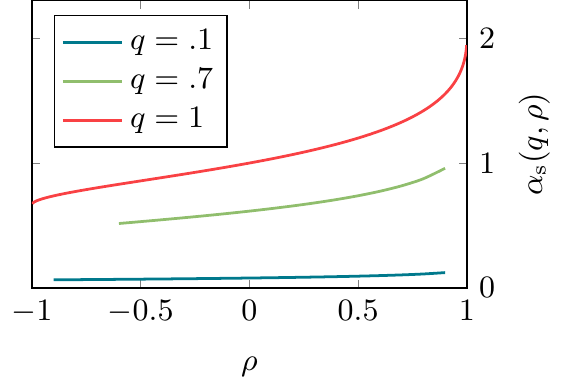}
\caption{Numerical solutions of the RS saddle-point equations~\eqref{eq:FRSB_RS_saddle} for the volume $V_{\text{s}}$ in~\eqref{eq:volume_LSO}. Left: $q$ as a function of $\alpha$ for different values of $\rho$; right: $\alpha$ at a fixed $q$ as a function of $\rho$. The curve $\alpha(1,\rho)$ is obtained from the asymptotic~\eqref{eq:RS_LSO_alphaC}, the other curves have limited supports due to numerical instability. Note that $\alpha(q,\rho)\to 0$ for $q\to 0$.}
\label{fig:saddleG}
\end{figure}

\begin{figure}[t]
\centering
\includegraphics[scale=1]{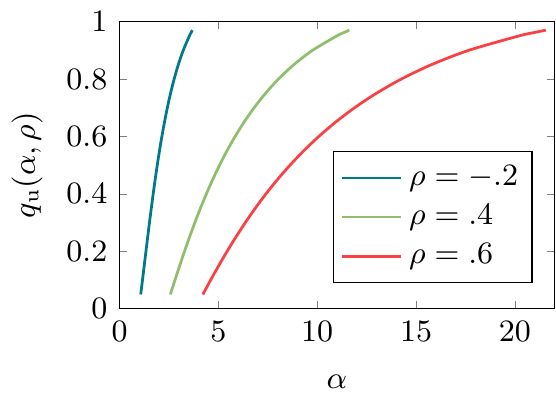}
\hspace{2em}
\includegraphics[scale=1]{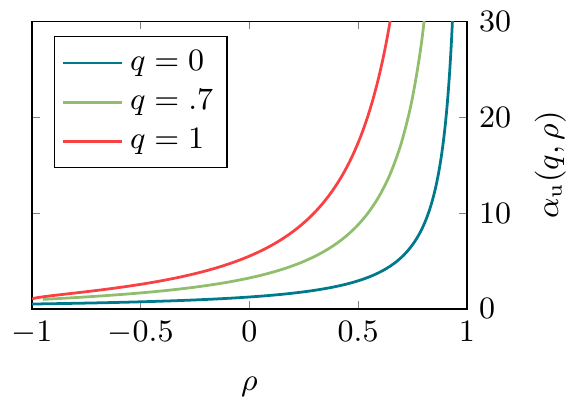}
\caption{Numerical solutions of the RS saddle-point equations~\eqref{eq:FRSB_RS_saddle} for the volume $V_\text{u}$ in~\eqref{eq:volume_GPR}. Left: $q$ as a function of $\alpha$ for different values of $\rho$; right: $\alpha$ at a fixed $q$ as a function of $\rho$. The curves $\alpha(1,\rho)$ and $\alpha(0,\rho)$ are obtained, respectively, from the asymptotic~\eqref{eq:RS_GPR_alphaStar} and from~\eqref{eq:FRSB_RS_GPRdAT}.}
\label{fig:saddleGPR}
\end{figure}

To write the RS free energy from Eq.~\eqref{eq:FRSB_functional}, I can simply use $q(x)=q_M=q_m$ constant and $f(q,\bm{h}) = f(q_M,\bm{h})$, to get
\begin{equation}
S^\text{RS}(q_M) = \frac{1}{2}\left[ \log(1-q_M)+\frac{q_M}{1-q_M}\right] +\alpha \left.\gamma_{q_M} \!\star\! f(q_M,\bm{h})\right|_{\bm{h}=0}\,.
\end{equation}
Notice that $f(q_M,\bm{h})$ is defined as in the starting condition of the Parisi equation~\eqref{eq:FRSB_Parisi}. The RS saddle point to evaluate the optimal value of $q_M$ is
\begin{equation}
\frac{q_M}{(1-q_M)^2} = \alpha \int \frac{\Diff2 \bm{h}}{2\pi q_{M} \sqrt{\det \mathcal R}} e^{-\frac{1}{2q_M} \bm{h}^T \mathcal{R}^{-1}\bm{h}}\left[\bm{\partial}f(q_M,\bm{h})\right]^T\mathcal{R}\left[\bm{\partial}f(q_M,\bm{h})\right]\,.
\label{eq:FRSB_RS_saddle}
\end{equation}
This equation can be solved for $q_M$ for any value of $\rho$, $\alpha$ or, equivalently, for $\alpha$ fixing $\rho$, $q_M$. Fig.~\ref{fig:saddleG} and \ref{fig:saddleGPR} report the numerical solutions of these saddle-point equations for the volumes, respectively, \eqref{eq:volume_LSO} and \eqref{eq:volume_GPR}. The limiting values of $\alpha$ at the saddle point for $q_M\to 1$ can be worked out analytically (see~\cite{Borra:2019,GPR:2020PRE}); they are, respectively,
\begin{align}
\alpha_c^{\text{RS}} (\rho) \equiv \alpha_{\text{s}}^{\text{RS}} (1,\rho) &= \left[\frac{2}{\pi}\arctan\!\left(\sqrt{\frac{1-\rho}{1+\rho}}\right)+\frac{1}{2} \right]^{-1}\,,
\label{eq:RS_LSO_alphaC}\\
\alpha_*^{\text{RS}} (\rho) \equiv \alpha_{\text{u}}^{\text{RS}} (1,\rho) &= \left[\frac{2}{\pi} \arctan\!\left(\sqrt{\frac{1-\rho}{1+\rho}}\right) -\frac{\sqrt{1-\rho^2}}{\pi}\right]^{-1}\,.
\label{eq:RS_GPR_alphaStar}
\end{align}
These curves correspond to the SAT/UNSAT transition lines for the two CSPs defined in Sec.~\ref{sec:volumes}, in the RS approximation. Indeed, when the satisfiability phase transition of a certain problem is approached from the SAT phase, the number of its solutions decreases, their volume in the space of the weights shrinks, so that different replicas of the systems become more and more correlated; as can be seen from the definition of the replica matrix, exactly at the transition $q_M\to 1$.


Notice that $\alpha_c$ is the analogous of the storage capacity found by Gardner for this kind of dataset: indeed, for $\rho\to 1$ the points in a doublet are coincident and $\alpha_c \to 2$, the Gardner's value for isolated points (see Sec.~\ref{sec:margin} and references therein). On the other hand, $\alpha_*$ diverges in the same limit: it is always possible to linearly classify a set of isolated points, if their labels can be chosen freely.

However, in the working hypothesis that a transition to a full RSB solution is present, also Eq.~\eqref{eq:FRSB_saddle2}, which has been obtained using explicitly the constraints from the Parisi equation, has to hold in the phase where $q(x)$ is a continuous function. Near the RS, it reads
\begin{equation}
\frac{1}{(1-q_M)^2} = \alpha \int \Diff2 \bm{h} \,\gamma_{q_M}(\bm{h}) \Tr\!\left[\mathcal{H}^{\text{RS}}(q_M,\bm{h})\mathcal{R}\mathcal{H}^{\text{RS}}(q_M,\bm{h})\mathcal{R}\right]\,,
\label{eq:FRSB_RS_stability}
\end{equation}
where
\begin{equation}
\mathcal{H}^{\text{RS}}(q_M,\bm{h}) = \begin{pmatrix}
\partial_{hh}	& \partial_{h\bar{h}}\\
\partial_{h\bar{h}} & \partial_{\bar{h}\bar{h}} 
\end{pmatrix} f(q_M,\bm{h})
\end{equation}
and $q_M=q_M(\alpha,\rho)$ is a solution of Eq.~\eqref{eq:FRSB_RS_saddle}. This means that the quantity
\begin{equation}
r(\alpha,\rho) = 1 - \alpha (1-q_M)^2 \int \Diff2 \bm{h} \,\gamma_{q_M}(\bm{h}) \Tr\!\left[\mathcal{H}^{\text{RS}}(q_M,\bm{h})\mathcal{R}\mathcal{H}^{\text{RS}}(q_M,\bm{h})\mathcal{R}\right]\,,
\label{eq:FRSB_RS_replicon}
\end{equation}
evaluated at the saddle point, must be null at the transition. Notice that to define $r(\alpha,\rho)$ I have rescaled Eq.~\eqref{eq:FRSB_RS_stability} by a factor $(1-q_M)^2$, to get rid of the trivial divergent behavior for $q_M\to 1$. This quantity is called \emph{replicon eigenvalue}: it is well known, indeed, that $r(\alpha,\rho)/(1-q_M)^2$ corresponds to (minus) the dangerous eigenvalue of the Hessian matrix $\partial^2 S/\partial Q_{cd} \partial Q_{ab}$ evaluated at the RS saddle point, and has to be positive if the RS ansatz is a stable extremal point of the action~\eqref{eq:S}. The line in the phase space where $r=0$, and so where the RS solution becomes unstable, is usually called de Almeida-Thouless (dAT) line, from~\cite{dAT:1978}. The fact that the de Almeida-Thouless stability analysis can be re-obtained from a full RSB scheme is not a coincidence: indeed, the Parisi continuous RSB solution is known to be \emph{marginally stable}, which is exactly the physical content of Eq.~\eqref{eq:FRSB_saddle2}. I will postpone a discussion on this crucial point to Sec.~\ref{sec:end}.

As the two problems studied in this paper present very different critical properties, I prefer to split the stability analysis in the next paragraphs, to deal with them separately and to ease the discussion.

\subsection{Supervised problem: marginal stability of the RS solution}
\label{sec:LSO}

\begin{figure}
\centering
\includegraphics[scale=1]{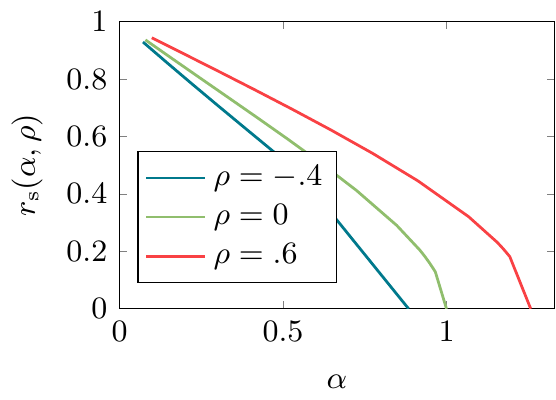}
\hspace{2em}
\includegraphics[scale=1]{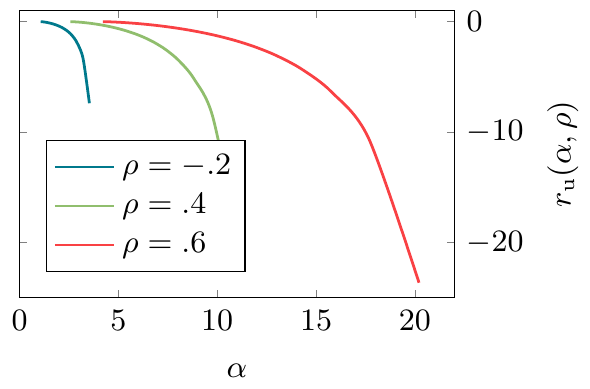}
\caption{Value of the replicon eigenvalue $r$ in Eq.~\eqref{eq:FRSB_RS_replicon} as a function of $\alpha$ at the RS saddle point for different values of $\rho$, for the volume $V_\text{s}$ in~\eqref{eq:volume_LSO} (left) and for the volume $V_{\text{u}}$ in~\eqref{eq:volume_GPR} (right). While in the first case $r$ is always positive and the RS solution is stable, in the second $r$ is negative for any values of $\alpha$ but $\alpha(q_M=0)$, when it is zero. For the supervised case, $r(\alpha,\rho)$ is exactly zero at $\alpha_c^\text{RS}(\rho)$ (the end point for each line is evaluated via the asymptotic analysis reported in App.~\ref{app:stabilityLSO}).}
\label{fig:replicon}
\end{figure}

In the case of the sCSP volume~\eqref{eq:volume_LSO}, the replicon is always positive and the RS ansatz is stable up to the satisfiability critical line. I report the numerical results in Fig.~\ref{fig:replicon}, left. Moreover, in the asymptotic limit $q_M\to 1$, that is evaluating Eq.~\eqref{eq:FRSB_RS_replicon} at $\alpha=\alpha_c^{\text{RS}} (\rho)$, it can be shown (see App.~\ref{app:stabilityLSO}) that $r$ is exactly zero for all values of $\rho$: this means that the RS solution for this problem is marginally stable at the SAT/UNSAT transition, that is
\begin{equation}
\alpha_{\text{s}}^{\text{dAT}}(\rho) = \alpha_{c}^{\text{RS}}(\rho)\,.
\label{eq:FRSB_RS_LSOdAT}
\end{equation}

The reason for this can be easily understood with simple arguments. For $\rho = 1$, for example, the problem reduces to the classification of $p$ isolated points with no structure, with storage capacity $\alpha_c^{\text{RS}} (1) = 2$: in this case, which is the one studied by Gardner at zero margin, it has been known since~\cite{Gardner:1988b} that the RS solution is marginally stable; for $\rho = 0$, the points in a doublet are essentially uncorrelated, so the problem is equivalent to the classification of $2p$ isolated points, with storage capacity $\alpha_c^{\text{RS}} (1) = 1$ (because $2p_c/n = 2$) and the same property of marginal stability; for other values of $\rho$ similar arguments can be formulated.  In other words, the problem with correlated patterns can always be mapped into an isolated-points problem with an effective number of patterns accounting for the amount of information present at the beginning.

The above reasoning should be enough to understand the difference between the overlap $\rho$ in the sCSP and the margin $\kappa$ of the usual Gardner problem, which corresponds to the classification of spheres with radius $\kappa$. In that case, any $\kappa>0$ is enough to make the transition non-critical (stable with a strictly positive replicon). I can say that there, because of the isotropy of the problem, all the directions in $\mathbb{R}^n$ are affected, while in the present case the fact that the points in a multiplet have a finite overlap means that the object is not a point only for a certain number of directions, which remains finite for $n\to\infty$, and the transition is critical as in the case of structureless points. I will clarify the parallelism between the two problems in Sec.~\ref{sec:margin}.

\subsection{Unsupervised problem: full Replica Symmetry Breaking}
\label{sec:GPR}

For the volume $V_\text{u}$ in~\eqref{eq:volume_GPR}, the replicon is always negative except for $q_M = 0$. This means that the dAT line is the $\alpha(q_M =0)$ line reported in Fig.~\ref{fig:saddleGPR} (right). This limit can be obtained analytically from Eq.~\eqref{eq:FRSB_RS_saddle}, with the change of variables $\bm{h}\to \sqrt{q_M}\bm{h}$ and expanding the RHS for small $q_M$; see App.~\ref{app:GPR} for details. The result is
\begin{equation}
\alpha_{\text{u}}^{\text{dAT}}(\rho) = \alpha_{\text{u}}^{\text{RS}}(0,\rho) = \frac{\left[2 \arcsin(\rho )+\pi \right]^2}{8 \left(1-\rho ^2\right)}\,.
\label{eq:FRSB_RS_GPRdAT}
\end{equation}
In the plane $\rho$--$\alpha$, this curve is always below the SAT/UNSAT critical line in the RS ansatz, Eq.~\eqref{eq:RS_GPR_alphaStar}, meaning that the true transition line for this problem has to be evaluated in an RSB approach.

Moreover, at variance with the previous case, where $\alpha_{\text{s}}^{\text{RS}}(q_M,\rho) \to 0$ for $q_M \to 0$, the line $\alpha_{\text{u}}^{\text{RS}}(0,\rho)$ is always different from zero. This means that for values of $\alpha$ below this curve, there is no solution of the RS saddle point equation for $q_M$ in the interval $[0,1]$, the free energy $S^{\text{RS}}(q_M)$ is a monotonic function in $q_M$ and the minimum is in $q_M = 0$.\footnote{In principle, solutions of the RS saddle point equation must be sought for $q_M$ in the whole interval $[-1,1]$. However, I could not find any real solution for $q_M$ negative; notice indeed that in this case the convolution~\eqref{eq:convolution} is ill defined, and a complex Hubbard-Stratonovich transformation in the original action~\eqref{eq:S} produces a complex $f$.}

\begin{figure}[t]
\centering
\includegraphics[height=12em]{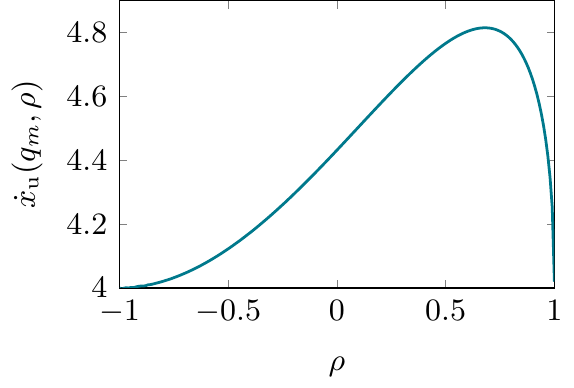}
\hspace{2em}
\includegraphics[height=12em]{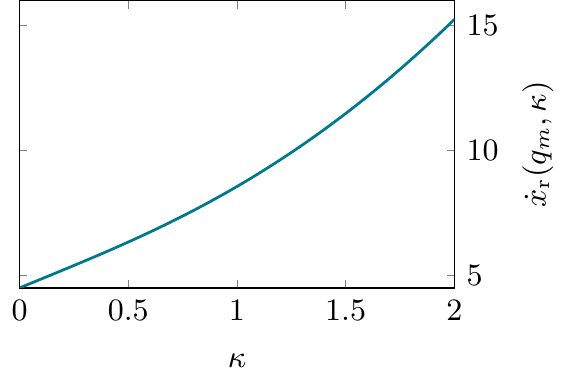}
\caption{Slope of the function $x(q)$ at the breaking point $q_m$ of the RS ansatz: left, in the case of the unsupervised problem as a function of the doublet $\rho$, Eq.~\eqref{eq:FRSB_RS_GPRslope2}; right, in the case of the resummed Gardner problem discussed in Sec.~\ref{sec:margin}, as a function of the margin $\kappa$, Eq.~\eqref{eq:margin_slope}. In both cases, $q_m=0$, $x(q_m)=0$.}
\label{fig:slope}
\end{figure}

The breaking point of the RS ansatz can be evaluated from Eq.~\eqref{eq:FRSB_saddle3}, which reads
\begin{equation}
x_m = \frac{1-q_M}{2} \frac{\int \Diff2 \bm{h}\, \gamma_{q_M}(\bm{h}) [\partial_\alpha \partial_\beta\partial_\gamma f(q_M,\bm{h})]\mathcal{R}_{\alpha\alpha'}\mathcal{R}_{\beta\beta'} \mathcal{R}_{\gamma\gamma'}[\partial_{\alpha'} \partial_{\beta'} \partial_{\gamma'} f(q_M,\bm{h})] }{\int \Diff2 \bm{h} \,\gamma_{q_M}(\bm{h})\left(
\Tr\{[\mathcal{H}(q_M,\bm{h})\mathcal{R}]^2\} + (1-q_M) \Tr\{[\mathcal{H}(q_M,\bm{h})\mathcal{R}]^3\}\right)}\,,
\end{equation}
to be evaluated at the instability line, that is for $q_M \to 0$. Note that for $q_M$ small the numerator is $O(q_M)$ (each derivative carries a factor $q_M^{-1/2}$, but I have to expand $f(q_M,\sqrt{q_M}\bm{h})$ up to order $q_M^2$ to get a non-null result from the third derivative), while the denominator is of order 1, so I get $x_m=0$. Given this information, the equation for the slope at the breaking point~\eqref{eq:FRSB_saddle4} becomes a lot easier:
\begin{equation}
\dot{x}(q_m) = \frac{\displaystyle \int \Diff2 \bm{h}\,\gamma_{q_M}(\bm{h}) [\partial^4_{\alpha\beta\gamma\delta} f(q_M,\bm{h} )]\mathcal{R}_{\alpha\alpha'}\mathcal{R}_{\beta\beta'} \mathcal{R}_{\gamma\gamma'}\mathcal{R}_{\delta\delta'}[\partial^4_{\alpha'\beta'\gamma'\delta'}f(q_M,\bm{h})]}{\displaystyle 2\int \Diff2 \bm{h} \,\gamma_{q_M}(\bm{h})\left(
\Tr\{[\mathcal{H}(q_M,\bm{h})\mathcal{R}]^2\} +  \Tr\{[\mathcal{H}(q_M,\bm{h})\mathcal{R}]^3\}\right)}\,,
\label{eq:FRSB_RS_GPRslope}
\end{equation}
keeping only the terms which remain different from 0 when $q_M \to 0$. I can evaluate its value analytically in this limit, obtaining
\begin{equation}
\dot{x}(q_m,\rho) = \frac{6 \pi  \rho\sqrt{1-\rho ^2}   L(0,0;\rho )-3 \rho ^2+3}{2 \pi ^2 L(0,0;\rho )^2}+2 \rho ^2+2\,,
\label{eq:FRSB_RS_GPRslope2}
\end{equation}
where the function $L$ is given in Eq.~\eqref{eq:L00}. The result is plotted in Fig.~\ref{fig:slope}, left: the slope is always positive for any value of $\rho$, meaning that the transition is to a full RSB phase. 

I will postpone considerations about these results and their relevance to Sec.~\ref{sec:end}. In the next section, I will perform the same stability analysis for the classification problem of another simple model of structured data, $(n-1)$-dimensional hyperspheres, pointing out the common features with the problems presented in this paper.

\section{Analogy and difference with the problem of learning with a margin}
\label{sec:margin}

In this section I will reconsider the classic task of learning with a margin from the point of view of data structure, elaborating on the concept of admissible classification; in doing so, I will make clear its connection with the CSPs I dealt with so far. An $n$-dimensional hyperplane described by the normal vector $W$ is a $\kappa$-margin classifier if it assigns the label $\sigma$ to the point $x\in \mathbb{R}^n$ according to the rule
\begin{equation}
\sigma = \begin{cases}
+ 1 & \text{if $\,\sum_{j=1}^n W_j x_j >\kappa \sqrt{n}$}\,,\\
- 1 & \text{if $\,\sum_{j=1}^n W_j x_j < -\kappa \sqrt{n}$}\,.
\end{cases}
\label{eq:margin_rule}
\end{equation}
Hence, $\kappa$ is the minimum distance between any point $x$ and the decision plane in order for the classification to be valid. Originally introduced to enhance the stability of learning algorithms, margin learning turned out to be crucial in the theory of support vector machines (SVMs), which are optimal margin classifiers.

\begin{figure}
\centering
\includegraphics[width=.8\textwidth]{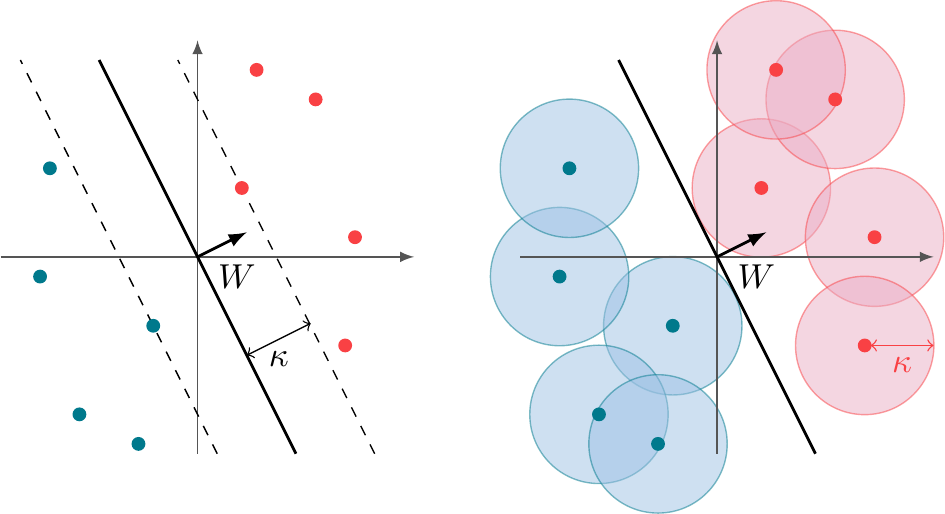}
\caption{Learning with a margin is equivalent to classifying spheres.}
\label{fig:spheres}
\end{figure}

Note that if a point $x$ falls closer than $\kappa$ to the plane $W$, its label is not defined by the rule~\eqref{eq:margin_rule}; using the jargon I adopted extensively in this paper, I can say that the classification realized by $W$ is \emph{not admissible} with respect to $x$. Moreover, I can push the analogy further: the problem of classifying points with a margin is indeed equivalent to the problem of classifying spheres with radius equal to the margin, as explained in detail in~\cite{Sompolinsky:2018prx}; see also Fig.~\ref{fig:spheres}.

The volume counting the admissible planes realizing a typical classification, that is those planes farther than $\kappa$ from any point in a typical instance of the training set, was introduced long ago by E. Gardner, in the seminal paper~\cite{Gardner:1988b}, for the case of points $x$ randomly chosen on the vertices of a $n$-dimensional hypercube (the cases of points on a hypersphere or i.i.d. Gaussian are equivalent, the only crucial ingredient being the vector $W$ normalized with a spherical, and not a discrete, constraint). The Gardner volume is defined as
\begin{equation}
V_{\text{G}} = \!\int\! \left[ \prod_{j=1}^n \diff W_j \right]\!\delta\! \left(\sum_{j=1}^n W_j^2 -n \right)
\!\prod_{\mu=1}^{p}  \theta\!\left(\frac{\sigma^\mu}{\sqrt{n}} \sum_{j=1}^n W_j x_{j}^\mu  - \kappa \right)
 \,. \label{eq:volume_G}
\end{equation}
Here, the instance of the labeled training set takes the role of a quenched disorder, as in the volume $V_\text{s}$ in~\eqref{eq:volume_LSO} (which was actually introduced later). I will not report here the stability analysis of this volume, which is well known from~\cite{Gardner:1988b} (see, for a careful discussion keeping into account also the case with $\kappa<0$,~\cite{Franz:2017scipost}). However, let me point out for reference that in the replica approach the RS ansatz is always stable (with a strictly positive Hessian) up to the SAT/UNSAT transition for any $\kappa>0$, and becomes marginally stable (the replicon eigenvalue becomes zero) only at $\kappa=0$. In this sense the margin $\kappa$, which is the parameter accounting for the structure of data in this problem, is different from the overlap $\rho$ of the sCSP: there, the RS ansatz remains marginal for any value of $\rho$. A possible hand-waving explanation for this difference is the fact that correlated points in a doublet are somewhat equivalent to a lower number of isolated points, because of the loss of information due to correlation. The Gardner's result for the critical capacity is
\begin{equation}
\alpha_c^{\text{RS}}(\kappa)= \left[\int_{-\kappa}^{+\infty} \frac{\diff h\, e^{-\frac{h^2}{2}} }{\sqrt{2 \pi }}\left(\kappa + h\right)^2 \right]^{-1}\,.
\end{equation}

What is more relevant in the context of this paper is the analogous of the volume $V_\text{u}$ in~\eqref{eq:volume_GPR} for margin learning. When the labels can be chosen freely, the only request on $W$ being the one of admissibility with respect to a typical configuration of the points, I can study the resummed volume~\cite{GPR:2020PRE}
\begin{equation}
\begin{aligned}
V_{\text{r}} &= \!\int\! \left[ \prod_{j=1}^n \diff W_j \right]\!\delta\! \left(\sum_{j=1}^n W_j^2 -n \right)
\!\prod_{\mu=1}^{p} \sum_{\sigma^\mu = \pm 1}\!\! \theta\!\left(\frac{\sigma^\mu}{\sqrt{n}} \sum_{j=1}^n W_j x_{j}^\mu  - \kappa \right)\\
&= \!\int\! \left[ \prod_{j=1}^n \diff W_j \right]\!\delta\! \left(\sum_{j=1}^n W_j^2 -n \right)
\!\prod_{\mu=1}^{p} \theta\!\left(\frac{1}{n} \sum_{i,j=1}^n W_i W_j x_i^\mu x_{j}^\mu  - \kappa^2 \right)
 \,, \label{eq:volume_kappa}
\end{aligned}
\end{equation}
which is obtained from~\eqref{eq:volume_G} summing over the labels. The second line of this equation should make clear the connection with~\eqref{eq:volume_GPR}: the $\kappa\to 0$ limit of $V_\text{r}$ should be equivalent to $V_{\text{u}}$ when $\rho\to 1$. However, both the limits are singular (the two problems become always trivially SAT), so this formal equivalence is misleading, as I will explain in the following.

In the replica approach, I can proceed as before, except that now there is no need to introduce a bidimensional gap variable. The variational free energy, obtained replicating and averaging the volume~\eqref{eq:volume_kappa}, can be written, in the RS approximation, as
\begin{equation}
S^\text{RS}_\text{r}(q_M) = \frac{1}{2}\left[ \log(1-q_M)+\frac{q_M}{1-q_M}\right] +\alpha \int \frac{\diff h \,e^{-\frac{h^2}{2q_M}} }{\sqrt{2\pi q_M}} f_\text{r}(q_M,h)\,,
\end{equation}
where the function $f_\text{r}$ is defined as
\begin{equation}
f_{\text{r}}(q_M,h) = \log\left[\frac{1}{2}\erfc\left(\frac{\kappa + h}{\sqrt{2(1-q_M)}}\right) + \frac{1}{2}\erfc\left(\frac{\kappa - h}{\sqrt{2(1-q_M)}}\right) \right]\,.
\end{equation}
The saddle point equation to find the optimal value of $q_M$ given $\alpha$ and $\kappa$ (or $\alpha$ given $q_M$ and $\kappa$) is
\begin{equation}
\frac{q_M}{(1-q_M)^2} = \alpha \int \frac{\diff h\,  e^{-\frac{h^2}{2q_M} }}{\sqrt{2\pi q_{M}}}\left[f_\text{r}'(q_M,h)\right]^2\,,
\label{eq:margin_RS_saddle}
\end{equation}
where the apex denotes a derivative with respect to $h$. The solution for $q_M\to 1$ was already reported in~\cite{GPR:2020PRE} and coincides with the critical value of $\alpha$ at the SAT/UNSAT transition of this problem in the RS approximation:
\begin{equation}
\alpha_*^\text{RS} (\kappa) = \frac{1}{2}\left[ \int_0^{\kappa } \frac{\diff h\, e^{-\frac{h^2}{2}} }{\sqrt{2 \pi }} (\kappa - h)^2\right]^{-1}\,.
\end{equation} 
As for the case of the limit $\rho \to 1$ in the uCSP, this quantity is diverging for $\kappa \to 0$, where no data structure is present.

The stability analysis follows closely the one in~\cite{Franz:2017scipost}. The dAT marginal stability condition is
\begin{equation}
\frac{1}{(1-q_M)^2} = \alpha\int \frac{\diff h \,e^{-\frac{h^2}{2 q_M}} }{\sqrt{2 \pi  q_M}}\left[f''_\text{r}(q_M,h)\right]^2\,,
\end{equation}
to be evaluated on the solution of the saddle point equation. As in the case discussed in Sec.~\ref{sec:GPR}, the only (marginally) stable solution is for $q_M = 0$, all the solutions for $q_M>0$ being unstable; the dAT line is
\begin{equation}
\alpha_\text{r}^\text{dAT} (\kappa) = \frac{\pi  \erfc\!\left(\kappa/\sqrt{2}\right)^2}{2 \kappa ^2 e^{-\kappa ^2}}\,.
\end{equation}
Again, the breaking point of the RS solution is in $x_m = 0$. 

An interesting difference with the uCSP case can be found in the evaluation of the slope of the order parameter function at the breaking point $x_m = 0$. The equation for the slope of the inverse function $x(q)$, analogous to Eq.~\eqref{eq:FRSB_saddle4} for a problem with a monodimensional gap variable, is
\begin{equation}
\dot{x}(q) = \frac{2 x(q)}{\alpha \int \diff h\,P(q,h) f'''(q,h)^2} \left[\frac{\alpha}{2} A(q)  - \frac{3 x(q)^2}{\lambda(q)^4} \right]\,,
\end{equation}
with
\begin{equation}
A(q) = \int \diff h\, P(q,h) \left[f''''(q,h)^2 - 12 x(q) f''(q,h) f'''(q,h)^2  + 6 x(q)^2 f''(x,q)^4 \right]\,,
\end{equation}
which gives, on the RS solution of the present problem for small $q$,
\begin{equation}
\dot{x}_\text{r}(q_m,\kappa) = \frac{\left(\sqrt{2 \pi } e^{\frac{\kappa ^2}{2}} \left(\kappa ^2-3\right) \erfc\left(\kappa/\sqrt{2}\right)-6 \kappa \right)^2}{4 \pi  e^{\kappa ^2} \erfc\left(\kappa/\sqrt{2}\right)^2+4 \sqrt{2 \pi } e^{\frac{\kappa ^2}{2}} \kappa  \erfc\left(\kappa/\sqrt{2}\right)}\,.
\label{eq:margin_slope}
\end{equation}
This result is always positive and monotonically increasing in $\kappa$, showing thus that even in this case the transition is to a full RSB phase; see Fig.~\ref{fig:slope}, right. A part from that, an interesting fact can be seen in the limit $\kappa \to 0$. Indeed,
\begin{equation}
\lim_{\kappa \to 0} \dot{x}_\text{r}(q_m,\kappa) = \frac{9}{2}\,,
\end{equation}
which is different from the limit $\rho \to 1$ of the equation for the slope in the uCSP, Eq.~\eqref{eq:FRSB_RS_GPRslope},
\begin{equation}
\lim_{\rho \to 1} \dot{x}_\text{u}(q_m,\rho) = 4 \,.
\end{equation}
In this sense, it seems that the two problems approach in different ways the limit where they become trivially SAT. In other words, starting from a problem where both the doublet structure and the margin are present, the limits $\rho\to 1$, $\kappa \to 0$ do not commute. It is still not clear to me why this is the case.

In the following section, I reconsider the classification problem of doublets, applying the results I reported so far from a point of view very promising for applications: the semi-supervised learning task of multi-view data, which I introduced briefly in Sec.~\ref{sec:intro}.

\section{Semi-supervised approach to the learning problem of structured data}
\label{sec:semisup}

As should be clear from the discussion so far, the unsupervised approach to learn structured data is considerably harder than the supervised one: first of all, it corresponds to a non-convex optimization problem with a lot of disconnected clusters of solutions; moreover, in the same way as the storage capacities of the two problems are such that $\alpha_*\gg \alpha_c$, also the number of unlabeled examples needed for an architecture trained only with the unsupervised approach to reach a low generalization error is expected to be much higher than the one needed in the supervised setting. 

This last claim is quite natural: if a solution (a separating hyperplane) does exist, a lot of unlabeled doublets are needed in order to exclude the rest of the synaptic volume, even in the over-optimistic hope that the doublets would fill homogeneously all the space except the portion where the plane is passing. A more precise formulation of the generalization problem in the statistical physics approach requires an analysis that goes beyond the scope of the present paper and is left for future investigation.

However, in many realistic scenarios where each point of the training set comes quite naturally in the form of different ``views'' of the same object~\cite{Zhao:2017}, it may become profitable to adopt a mixed strategy. Think for example of the task of classifying clips from a movie with labels certifying if a certain actor is or is not on the screen. Each clip comes with different views (a video stream, an audio track) that correspond to the same label (yes or no). While forming an unlabeled multi-view dataset in this setting is straightforward, a supervised-learning training set requires a costly human contribution to label the patterns. 

\begin{figure}
\centering
\includegraphics[width=.4\textwidth]{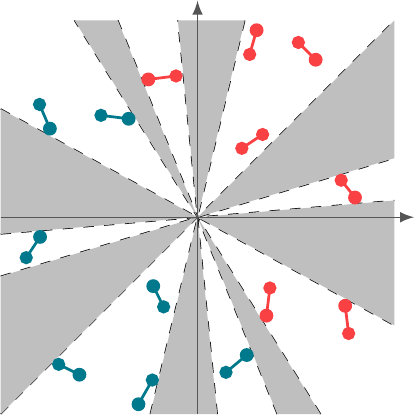}
\caption{Linear classification of doublets as a semi-supervised task. The planes spanning any grey region are solutions of the unsupervised problem~\ref{problem:GPR}, which ignores the labels in the training set and only searches for a classification not breaking any segment; a lot of patterns are needed in order to obtain the solutions with good generalization properties, and the problem is hard due to ergodicity breaking. On the other hand, only the planes spanning the narrower region solve the supervised problem~\ref{problem:LSO} with the labeling depicted in figure, which is an easy task. In a setting where the labeling of the training set is costly, one should search for a trade-off between the two approaches using a mixed strategy.}
\label{fig:semisup}
\end{figure}

Hence, semi-supervised learning strategies have proven very successful: in co-training~\cite{Blum:1998}, two classifiers are trained on the two different views of the few clips in the labeled set, and then are used to cross-label clips from the unlabeled set. This protocol is performed in such a way to maximize the compatibility of the two classifiers, that is the fact that they must agree in labeling the views of the same object. In the geometrical toy-model analogy of multi-view learning I presented in this paper, the two views are the points in a doublet, and the principle of compatibility is the admissibility constraint on the classification; see Fig.~\ref{fig:semisup}.

Following this research line, I can say that the two CSPs studied here are the extreme cases (fully-labeled vs. fully-unlabeled dataset) of a more general semi-supervised problem with $\alpha_\text{s} n$ labeled examples and $\alpha_{\text{u}} n$ unlabeled ones, whose phase diagram can be studied in the same spirit: a synaptic volume counting the number of planes complying with the two sets of constraints can be defined. As the two datasets can be taken as independent, the variational action of this replicated averaged \emph{semi-supervised} volume can be easily obtained form the result I already reported; in the case of doublets, the RS approximation gives
\begin{equation}
S^\text{RS}_{\text{ss}}(q) = \frac{1}{2}\left[ \log(1-q)+\frac{q}{1-q}\right] +\alpha_\text{s} \left.\gamma_{q} \!\star\! f_\text{s}(q,\bm{h})\right|_{\bm{h}=0}+ \alpha_\text{u} \left.\gamma_{q} \!\star\! f_\text{u}(q,\bm{h})\right|_{\bm{h}=0}\,,
\end{equation}
where $f_\text{s}(q,\bm{h})$ and $f_\text{u}(q,\bm{h})$ are written, respectively, in Eq.~\eqref{eq:app_f_Gardner} and~\eqref{eq:app_f_GPR}.

\begin{figure}
\centering
\includegraphics[scale=1.1]{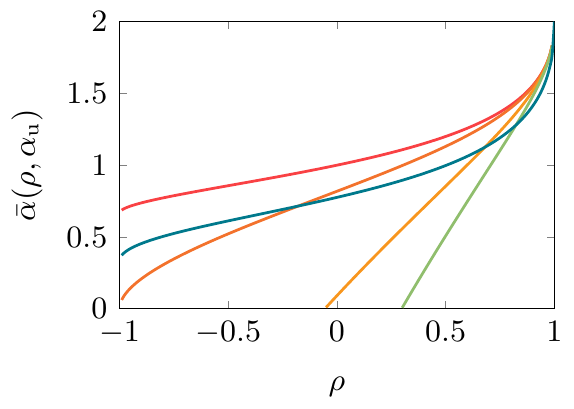}
\caption{Storage capacity $\bar{\alpha}$ of the semi-supervised problem, Eq.~\eqref{eq:alphaSS}, for different values of $\alpha_\text{u}$, quantifying the number of unlabeled patterns in the training set. Upper curve (red): $\alpha_\text{u}=0$, $\bar{\alpha}=\alpha_c$; orange, yellow, green curves: constant $\alpha_\text{u} = 1,\,5,\,10$ respectively; blue curve: $\alpha_\text{u} = \alpha_\text{u}^\text{dAT}(\rho)$ from Eq.~\eqref{eq:FRSB_RS_GPRdAT}, to plot a non-constant value of $\alpha_\text{u}$ in a region where the pure uCSP is still replica-symmetric. The inclusion of unlabeled constraints in the supervised problem lowers the critical capacity, except for $\rho\to 1$, where the doublet structure is lost.}
\label{fig:semisup_alpha}
\end{figure}

It is natural then to ask what is the critical number of labeled examples in order to reach the storage capacity limit $q\to 1$, \emph{at a fixed number of unlabeled examples}. The result follows from the asymptotics reported in App.~\ref{app:stability}; the critical value of $\alpha_\text{s}$, in the RS approximation, is
\begin{equation}
\bar{\alpha}(\rho,\alpha_\text{u}) = \alpha_c(\rho) \left[1 - \frac{\alpha_{\text{u}}}{\alpha_*(\rho)} \right]\,, \qquad \alpha_\text{u} < \alpha_*(\rho)\,,
\label{eq:alphaSS}
\end{equation}
where $\alpha_c$ and $\alpha_*$ are given, respectively, in Eq.~\eqref{eq:RS_LSO_alphaC} and~\eqref{eq:RS_GPR_alphaStar}. This result is plotted in Fig.~\ref{fig:semisup_alpha} for different values of $\alpha_\text{u}$: as expected, the inclusion in the problem of the additional constraints, coming from the admissibility condition on the unlabeled examples, reduces the storage capacity of the pure supervised one; $\bar{\alpha} \to 0$ when $\alpha_\text{u} \to \alpha_*$, the limit of capacity for the pure uCSP.

Though not reproducing the stability analysis of the previous sections, I expect the RS ansatz for this semi-supervised problem to be unstable in general, due to the inclusion of the non-linear constraints from the unlabeled dataset. However, I do not expect this qualitative picture to change significantly in a full RSB approach.

\section{Discussion and outlooks}
\label{sec:end}

In this paper I studied the critical properties of the SAT/UNSAT transitions occurring in different CSPs arising from the problem of learning structured data. In a simple model of input-label correlation in which the patterns to classify are doublets of points sharing the same label, one can think to solve the task of searching for a separating hyperplane, enforcing a linear decision rule, with two different strategies.

The first one consists in a supervised learning approach, where both the fact that the patterns in the training set come as inseparable doublets and the fact that they are labeled are exploited to infer the solution. The associated supervised CSP~\ref{problem:LSO} can be studied in a replica approach, searching for the number of solutions for a typical instance of a random-labeled training set. As I showed in Sec.~\ref{sec:LSO}, replica symmetry is exact up to the satisfiability transition. This means that the space of solutions is connected in a unique ergodic component, which shrinks with the number of patterns up to the SAT/UNSAT line. 

The second approach is blind to the label assignment of the training set: the only information used to infer the solution is the fact that the classification must be admissible, \textit{i.e.} it must be compatible with the structure of the data. The corresponding unsupervised CSP~\ref{problem:GPR}, formulated in~\cite{GPR:2020PRE}, is quadratically constrained and presents a drastic form of full RSB, as exlpained in Sec.~\ref{sec:GPR}: the satisfiability transition, which in this case is produced by the fact that the excluded volume due to data structure tends to fill the whole space of solutions when the number of patterns is increased, is preceded by an ergodicity-breaking phase transition, at which this space shatters into disconnected clusters. Moreover, the order parameter function $q(x)$, quantifying the overlap between different solutions of the problem, is a continuous function starting from the origin with a non-trivial slope; this result is rationalized saying that each of these disconnected components, which is a meta-basin of attraction in the corresponding optimization problem, is in turn broken into inner basins, and so forth in a continuous, hierarchical way.

In both cases, marginal stability is found at the satisfiability transition. This property is a feature of the continuous RSB transition that occurs, as I showed in this paper, both in the sCSP (at the satisfiability line) and in the uCSP (in the SAT phase), and its physical consequences have been extensively investigated in the spin glass and jamming literature. In  particular, when the SAT/UNSAT transition takes place in a marginally stable region, it has been included in the same universality class of isostatic jamming~\cite{FranzParisi:2016}: the number of ``contacts'', that is the number of constrains that are only marginally satisfied at the transition, is equal to the number $n$ of degrees of freedom in the problem; this is the analogous of the Maxwell stability criterion in mechanical systems. Other critical properties, such as the power law distribution of small positive or negative gaps near the transition (see, for example,~\cite{Wyart:2012,Franz:2017scipost}) can be related to this fact; far from giving an exhaustive review of the literature on the subject, I address the interested reader to~\cite{PUZ:2020} for reference. Here, it is enough to stress that this type of critical transition is qualitatively different from the so-called Random First Order Transition that is thought to be relevant in glassy dynamics: while the latter is associated to optimization problems where the insurgence of an exponential number of metastable states makes the global minima impossible to reach dynamically, the former is related to problems whose global optimal solutions are typically \emph{easy to approximate}, in the sense explored in~\cite{Montanari:2020}.

The point of view I adopted in this paper is that the current analysis can be particularly relevant in the context of semi-supervised learning of multi-view data. I showed indeed in Sec.~\ref{sec:semisup} how the storage capacity of a linear classifier, defined as the maximum number of randomly labeled objects an hyperplane can separate, is lowered by the inclusion of unlabeled examples in the dataset, demanding compatibility. 
A natural and promising follow up of the present work, which is focused exclusively on the problem of capacity, is to apply the same tools from statistical mechanics to the problem of generalization. 

Suppose indeed to be in a more realistic scenario where an optimal hyperplane separating the patterns does exist, because for example the labels of the training set are planted in the dataset by a teacher, see~\cite{EVDB:2001}. 
Note that in the case of structured data it is not so clear how to define a teacher providing labels: indeed, the teacher should respect the admissibility constraints, so any definition should be dependent on the instance of the dataset. A possible way could be decimating a randomly generated dataset in order to get rid of the doublets not compatible with the plan provided by the teacher. This procedure is obviously expected to stop working at $\alpha_*(\rho)$, and approaching this limit the teacher is more and more correlated to the distribution of the patterns. This difficulty aside, a natural question would be to search for an optimal ratio of labeled/unlabeled patterns that a semi-supervised algorithm needs to reach a low generalization error. PAC-Bayes bounds on the generalization error in this setting are rigorously proven only under more or less strict hypothesis on the distribution of the data (in co-training, the different views have to be conditional independent, or weakly dependent, given the label, see~\cite{Dasgupta:2002,Balcan:2005}), while a physical approach should be more flexible, though less rigorous.

Moreover, I can think of at least two ways to extend numerically the analysis presented so far. The first one consists in a numerical analysis of the phase diagram of the two CSPs, following for example~\cite{Sclocchi:2019}: an Hamiltonian can be defined for the vector $W$, weighting the violated constraints, whose zero minima coincide with the solutions of the problems. The critical properties presented here should corresponds to dynamical features of the algorithms used to solve the associated optimization problems. The second direction is the search for a numerical solution of the Parisi equation~\eqref{eq:FRSB_Parisi} in the unsupervised case, in order to find the true critical $\alpha_*$ line in a full RSB approach. This result, which is in other problems of limited practical interest, would be particularly intriguing here: indeed, in~\cite{GPR:2020PRE} a different combinatorial argument is formulated to obtain the (conjecturally) same curve. That approach is based on an uncontrolled mean-field approximation, which could be clarified in the light of the replica result.

\section{Acknowledgement}
I am supported by a grant from the Simons Foundation (grant No. 454941, S. Franz). I would like to warmly thank Silvio Franz, Pietro Rotondo, Marco Gherardi, Fabian Aguirre for discussions and suggestions.

\appendix

\section{Replicated-averaged volume}
\label{app:replicas}
Starting from a volume of the form~\eqref{eq:volume_LSO} or~\eqref{eq:volume_GPR}, I replicate it $t$ times, obtaining
\begin{multline}
V^t =   \int \left[\prod_{a=1}^t \prod_{j=1}^n \diff W_j^a \right] \prod_{a=1}^t \delta \!\left[\sum_{j=1}^n (W_j^a)^2 -n \right]\\
\times\int \left[\prod_{a=1}^t \prod_{\mu=1}^p \Diff2 \bm{y}^\mu_a \right] \!\prod_{\mu=1}^{p}\delta^{(2)}\!\!\left(\bm{y}^\mu_a - \frac{\sum_j W_j^a \bm{\xi}^\mu_j}{\sqrt{n}}\right)v\!\left(\bm{y}^\mu_a\right)\,,
\end{multline}
where $\delta^{(2)}(\bm{x}) = \delta(x)\delta(\bar{x})$ and $v$ has the form of the corresponding constraints. The average over the distribution of the patterns is performed by Fourier-transforming the $\delta^{(2)}$ and expanding to $O(1/n)$; the result depends only on the combination $Q_{ab} = \sum_j W^a_j W^b_j/n$. The integrals over the weights can be done introducing a delta enforcing this equality and integrating over $Q_{ab}$, to get Eq.~\eqref{eq:volume_diff}.

\section{Hierarchical RSB ansatz}
\label{app:hierarchicalRSB}

I report here for reference the details of the RSB hierarchical formulation of the problem using the formalism devised in~\cite{Duplantier:1981}. A $t\times t$ RS matrix can be written as
\begin{equation}
\begin{aligned}
Q^{\text{RS}}_t(1,q_0) &= (1-q_0) \mathbf{I}_t + q_0  \bm{1}_t\\
&=(1-q_0) \mathbf{I}_t \otimes \bm{1}_1  + q_0  \bm{1}_t\,,
\end{aligned}
\end{equation}
where $\mathbf{I}_t$ is the $t$-dimensional identity matrix, $\bm{1}_t$ the $t$-dimensional matrix with all entries equal to 1, $\otimes$ the Kronecker product. The hierarchy of $K$RSB matrices can be written iteratively, starting from 1RSB 
\begin{align}
&\begin{aligned}
Q^{\text{1RSB}}_{t,m_1}(1,q_1,q_0) 
&= \mathbf{I}_{t/m_1} \otimes Q^{\text{RS}}_{m_1}(1-q_0,q_1-q_0) + q_0 \bm{1}_t\,,\\
\end{aligned}
\end{align}
and so on. As long as $t$ is an integer, the order relation between the parameters defining these matrices is
\begin{equation}
\boxed{t\in \mathbb{N}^+}\qquad 1\le m_K \le\cdots \le m_1 \le t\,,\qquad 0\le q_0 \le \cdots\le q_K \le 1\,.
\end{equation}
As usual for the replica method, when $t\to 0$ the above order relation of the block-dimension parameters $m_i$ has to be reversed:
\begin{equation}
\boxed{t\in [0,1)}\qquad t\le m_1 \le\cdots \le m_K \le 1\,,\qquad 0\le q_0 \le \cdots\le q_K \le 1\,.
\end{equation}
When applied to a product function of $t$ replicas:
\begin{equation}
\begin{aligned}
&\left.e^{\frac{1}{2}\sum_{a,b} Q_{ab}^{\text{RS}} D_{ab}} v(\bm{h}_1)\cdots v(\bm{h}_t)\right|_{\bm{h}_a = 0} \\
&= \left. e^{\frac{1-q_0}{2}\sum_{a}D_{aa} + \frac{q_0}{2} \sum_{a,b}\bm{\partial}_a^T \mathcal{R} \bm{\partial}_b} v(\bm{h}_1)\cdots v(\bm{h}_t)\right|_{\bm{h}_a = 0}\\
&= \int \frac{\Diff2 \bm{z}}{2\pi q\sqrt{\det \mathcal{R}}}\left.  e^{-\frac{\bm{z}^T\mathcal{R}^{-1}\bm{z}}{2 q_0} + \bm{z} \cdot \bm{\partial} }\left[e^{\frac{1-q_0}{2} D} v(\bm{h})\right]^t\right|_{\bm{h} = 0}\,.
\end{aligned}
\end{equation}
I can write
\begin{equation}
\begin{aligned}
e^{\frac{1-q_0}{2} D} v(\bm{h})
&= \int \frac{\Diff2 \bm{k} }{ 2\pi(1- q_0 )\sqrt{\det{\mathcal{R}}}} v(\bm{k}+\bm{h}) e^{\frac{1}{2(1-q_0)} \bm{k}^T \mathcal{R}^{-1}\bm{k}}\\
&=\gamma_{1-q_0} \!\star\! v(\bm{h})\,,
\end{aligned}
\end{equation}
where the Gaussian convolution is defined by Eq.~\eqref{eq:convolution}. Finally,
\begin{equation}
\left.e^{\frac{1}{2}\sum_{a,b} Q_{ab}^{\text{RS}} D_{ab}} v(\bm{h}_1)\cdots v(\bm{h}_t)\right|_{\bm{h}_a = 0} = \left. \gamma_{q_0}\!\star\! \left[\gamma_{1-q_0} \!\star\! v(\bm{h}) \right]^t\right|_{\bm{h}=0}\,.
\end{equation}
Given this equation, I can write the above quantity for all the hierarchical matrices in an iterative way. For example,
\begin{equation}
\left.e^{\frac{1}{2}\sum_{a,b} Q_{ab}^{\text{1RSB}} D_{ab}} v(\bm{h}_1)\cdots v(\bm{h}_t)\right|_{\bm{h}_a = 0} = \left. \gamma_{q_0}\!\star\! \left\{\gamma_{q_1-q_0} \!\star\! \left[\gamma_{1-q_1}\!\star\! v(\bm{h})\right]^{m_1} \right\}^{t/m_1}\right|_{\bm{h}=0}\,.
\end{equation}
Recursively,
\begin{equation}
\begin{aligned}
g(1,\bm{h}) & = \gamma_{1-q_1}\!\star\! v(\bm{h})\\
g(m_1,\bm{h}) & = \gamma_{q_1-q_0} \!\star\! \left[g(1,\bm{h})\right]^{m_1}\\
G &= \left. \gamma_{q_0}\!\star\! \left[g(m_1,\bm{h})\right]^{t/m_1}\right|_{\bm{h}=0}\,.
\end{aligned}
\end{equation}
The $K$RSB ansatz is then defined by the recursion
\begin{equation}
\begin{aligned}
g(1,\bm{h}) & = \gamma_{1-q_K}\!\star\! v(\bm{h})\\
g(m_K,\bm{h}) & = \gamma_{q_K-q_{K-1}}\!\star\! \left[g(1,\bm{h})\right]^{m_K}\\
\cdots\\
g(m_i,\bm{h}) & = \gamma_{q_i-q_{i-1}}\!\star\! \left[g(m_{i+1},\bm{h})\right]^{m_i/m_{i+1}}\\
\cdots\\
g(m_1,\bm{h}) & = \gamma_{q_1-q_0} \!\star\! \left[g(m_2,\bm{h})\right]^{m_1/m_2}\\
G &= \left. \gamma_{q_0}\!\star\! \left[g(m_1,\bm{h})\right]^{t/m_1}\right|_{\bm{h}=0}\,.
\end{aligned}
\label{eq:RSBrec}
\end{equation}

\section{Continuous RSB}
I report here the details of the calculations needed to derive the equations in Sec.~\ref{sec:multiFullRSB}.

\label{app:FRSB}

\subsection{Parisi equation}
Following~\cite{Duplantier:1981,MPV:1986}, I define the stepwise function
\begin{equation}
q(x) = q_j \qquad \text{if}\,\,\, m_j \le x \le m_{j+1}\,.
\end{equation}
In the continuum limit of the RSB scheme,
\begin{equation}
m_j - m_{j+1} \to -\diff x\,.
\end{equation}
The generic term of the recursion~\eqref{eq:RSBrec} becomes
\begin{equation}
g(x-\diff x,\bm{h})  = \gamma_{\diff q}\!\star\! \left[g(x,\bm{h})\right]^{(x-\diff x)/x}\,.
\end{equation}
Using Eq.~\eqref{eq:convolution} and expanding to first order I get
\begin{equation}
\dot{g}(x,\bm{h}) = -\frac{1}{2} \frac{\diff q}{\diff x} \bm{\partial}^T\mathcal{R}\bm{\partial} g(x,\bm{h}) + \frac{g(x,\bm{h}) \log g(x,\bm{h})}{x}\,,
\end{equation}
with starting condition
\begin{equation}
g(1,\bm{h}) = g(x_M,\bm{h}) = \gamma_{1-q_M}\!\star\! v(\bm{h})\,.
\end{equation}
Now I can define
\begin{equation}
g(x,\bm{h}) = e^{x f(x,\bm{h})}
\end{equation}
to get
\begin{equation}
\left\{
\begin{aligned}
&f(1,\bm{h}) = \log \gamma_{1-q_M}\!\star\! v(\bm{h})\,,\\
&\dot{f}(x,\bm{h}) = -\frac{1}{2} \dot{q}(x)\left\{ \bm{\partial}^T\mathcal{R}\bm{\partial} f(x,\bm{h})  + x\left[\bm{\partial}f(x,\bm{h})\right]^T\mathcal{R}\left[\bm{\partial}f(x,\bm{h})\right] \right\}\,,\quad x_m\le x \le x_M\,.
\end{aligned}\right.
\end{equation}
In terms of the inverse function $x = x(q)$, defined for $q_m\le q \le q_M$, I obtain finally Eq.~\eqref{eq:FRSB_Parisi}. 

\subsection{Free energy functional}

I can now write the free energy functional~\eqref{eq:S} in the full RSB scheme. For the determinant part, I know that~\cite{MPV:1986}
\begin{equation}
\lim_{t\to 0} \frac{1}{t} \log \det Q = \log(1-q_M)+\frac{q_m}{\lambda(0)}+\int_0^1 \diff x \frac{\dot{q}(x)}{\lambda(x)} \,,
\end{equation}
with
\begin{equation}
\lambda(x) = 1 - x q(x) - \int_x^1\diff y \,q(y)\,.
\end{equation}
In terms of the inverse function $x(q)$,
\begin{equation}
\lim_{t\to 0} \frac{1}{t} \log \det Q = \log(1-q_M)+\frac{q_m}{\lambda(q_m)}+\int_{q_m}^{q_M} \frac{\diff q }{\lambda(q)}\,,
\end{equation}
where $\lambda(q) = \lambda[x(q)]$ is given by Eq.~\eqref{eq:FRSB_lambda}.
For the constraint part:
\begin{equation}
\begin{aligned}
\lim_{t\to 0} \frac{1}{t} \log \left.e^{\frac{1}{2} \sum_{a,b} Q_{ab}D_{ab}} \prod_a v(\bm{h}_a)\right|_{h_a=0} 
&= \lim_{\substack{t\to 0\\q_0\to q_m\\m_1\to x_m}} \frac{1}{t} \log \left\{ \left. \gamma_{q_0}\!\star\! \left[g(m_1,\bm{h})\right]^{t/m_1}\right|_{\bm{h}=0}\right\}\\
&=  \left. \gamma_{q_m}\!\star\! \frac{1}{x_m}\log\left[g(x_m,\bm{h})\right]\right|_{\bm{h}=0}\\
&= \left.\gamma_{q_m} \!\star\! f(0,\bm{h})\right|_{\bm{h}=0}
\end{aligned}
\end{equation}
where I used $f(x_m,\bm{h})=f(0,\bm{h})$. Summing the two parts I find
\begin{equation}
S[x(q)]= \frac{1}{2}\left[ \log(1-q_M)+\frac{q_m}{\lambda(q_m)}+\int_{q_m}^{q_M} \frac{\diff q }{\lambda(q)} \right] +\alpha \left.\gamma_{q_m} \!\star\! f(q_m,\bm{h})\right|_{\bm{h}=0}\,.
\end{equation}
Adding Lagrange multipliers to enforce the conditions~\eqref{eq:FRSB_Parisi}, I obtain the full functional~\eqref{eq:FRSB_functional} whose variations give $x(q)$ at the saddle point.

\subsection{Stability equation}
To get Eq.~\eqref{eq:FRSB_saddle2}, I derive Eq.~\eqref{eq:FRSB_saddle1} with respect to $q$:
\begin{equation}
\begin{aligned}
&\frac{\delta}{\delta q} \int \Diff2 \bm{h}\, P(q,\bm{h}) \left[\bm{\partial}f(q,\bm{h})\right]^T\mathcal{R}\left[\bm{\partial}f(q,\bm{h})\right]\\
&= \int \Diff2 \bm{h}\, \biggl\{ \frac{1}{2}(\bm{\partial}^T\mathcal{R}\bm{\partial} P)\left(\bm{\partial}f\right)^T\mathcal{R}\left(\bm{\partial}f\right)
+ (\bm{\partial}^T\mathcal{R}\bm{\partial} f)[\bm{\partial}^T\mathcal{R}(P\bm{\partial}f)]  \biggr\}\,,
\end{aligned}
\end{equation}
where I used the equations of motion for $f$ and $P$. The only tricky calculation is
\begin{equation}
\begin{aligned}
\frac{1}{2}\int \Diff2 \bm{h}\,  (\bm{\partial}^T\mathcal{R}\bm{\partial} P)\left(\bm{\partial}f\right)^T\mathcal{R}\left(\bm{\partial}f\right)
&=\frac{1}{2}\int \Diff2 \bm{h}\,  P\{\bm{\partial}^T\mathcal{R}\bm{\partial}[\left(\bm{\partial}f\right)^T\mathcal{R}\left(\bm{\partial}f\right)]\}\\
&=\mathcal{R}_{\alpha\beta}\mathcal{R}_{\gamma\delta}\int \Diff2 \bm{h}\,  P\{\partial_\alpha [\left(\partial_\beta\partial_\gamma f\right)\left(\partial_\delta f\right)]\}\\
&=\int \Diff2 \bm{h}\,  P\{ [\bm{\partial} (\bm{\partial}^T\mathcal{R}\bm{\partial} f)]^T\mathcal{R}\left(\bm{\partial} f\right)+\Tr[\mathcal{H}\mathcal{R}\mathcal{H}\mathcal{R}]\}
\end{aligned}
\end{equation}
where the Greek indices are summed. Assembling everything, I obtain Eq.~\eqref{eq:FRSB_saddle2}.

\subsection{Breaking point equation}
To get Eq.~\eqref{eq:FRSB_saddle3}, I derive with respect to $q$ Eq.~\eqref{eq:FRSB_saddle2}:
\begin{equation}
\frac{2 x}{\lambda^3} = \alpha \int \Diff2 \bm{h}\, (\dot{P} \Tr \mathcal{HRHR} + 2 P \Tr\dot{\mathcal{H}}\mathcal{RHR})
\label{eq:FRSB_saddle3_0}
\end{equation}
In the following I will omit for conciseness the integration over $\bm{h}$ and the arguments of the functionals. The first term on RHS is then (using the saddle point equations and integrating by parts):
\begin{multline}
(DP/2)\Tr (\mathcal{HRHR}) + x P(\bm{\partial}f)^T\mathcal{R}\bm{\partial}\Tr (\mathcal{HRHR}) \\
= P\Tr [(D\mathcal{H})\mathcal{RHR}] +  P (\partial_\alpha \partial_\beta\partial_\gamma f)\mathcal{R}_{\alpha\alpha'}\mathcal{R}_{\beta\beta'} \mathcal{R}_{\gamma\gamma'}(\partial_{\alpha'} \partial_{\beta'} \partial_{\gamma'} f) \\
 +x P(\bm{\partial}f)^T\mathcal{R}\bm{\partial}\Tr (\mathcal{HRHR})\,.
\end{multline}
Second term on RHS:
\begin{equation}
\begin{aligned}
&-P \Tr[(D\mathcal{H})\mathcal{RHR}] - x P \partial_\alpha \partial_\beta [(\partial_\kappa f) \mathcal{R}_{\kappa\lambda} (\partial_\lambda f)] \mathcal{R}_{\beta\gamma}(\partial_\gamma \partial_\delta f )\mathcal{R}_{\delta\alpha}\\
&=-P \Tr[(D\mathcal{H})\mathcal{RHR}] - x P (\bm{\partial}f)^T\mathcal{R}\bm{\partial}\Tr (\mathcal{HRHR}) - 2 x P \Tr (\mathcal{HRHRHR})\,.
\end{aligned}
\end{equation}
Summing and using again Eq.~\eqref{eq:FRSB_saddle2}, I obtain Eq.~\eqref{eq:FRSB_saddle3}.

\subsection{Slope equation}
To get the equation for the slope of the curve $q(x)$, I derive Eq.~\eqref{eq:FRSB_saddle3_0} with respect to $q$, to get
\begin{equation}
\begin{aligned}
\frac{2\dot{x}}{\lambda^3} - \frac{6x \dot{\lambda}}{\lambda^4} &= \alpha \int \Diff2 \bm{h} \Bigl[\dot{P}(\partial^3_{\alpha\beta\gamma}f )\mathcal{R}_{\alpha\alpha'}\mathcal{R}_{\beta\beta'} \mathcal{R}_{\gamma\gamma'}(\partial^3_{\alpha'\beta'\gamma'}f)\\
&\hphantom{= \alpha \int \Diff2 \bm{h} \Bigl[} +2P(\partial^3_{\alpha\beta\gamma}\dot{f} )\mathcal{R}_{\alpha\alpha'}\mathcal{R}_{\beta\beta'} \mathcal{R}_{\gamma\gamma'}(\partial^3_{\alpha'\beta'\gamma'}f)\\
&\hphantom{= \alpha \int \Diff2 \bm{h} \Bigl[} -2\dot{x}P \Tr(\mathcal{H}\mathcal{R})^3
-2x\dot{P} \Tr(\mathcal{H}\mathcal{R})^3
-6xP\Tr(\dot{\mathcal{H}}\mathcal{R}\mathcal{H}\mathcal{R}\mathcal{H}\mathcal{R})\Bigr]\\
&+\dot{\alpha} \int \Diff2 \bm{h} P \left(\partial^3 f \partial^3 f  - 2 x \Tr(\mathcal{HR})^3 \right)\,.
\end{aligned}
\end{equation}
\begin{equation}
\begin{aligned}
\text{RHS, first term}
&= P(\partial^3_{\alpha\beta\gamma} D f )\mathcal{R}_{\alpha\alpha'}\mathcal{R}_{\beta\beta'} \mathcal{R}_{\gamma\gamma'}(\partial^3_{\alpha'\beta'\gamma'}f) \\
&\quad+P(\partial^4_{\alpha\beta\gamma\delta} f )\mathcal{R}_{\alpha\alpha'}\mathcal{R}_{\beta\beta'} \mathcal{R}_{\gamma\gamma'}\mathcal{R}_{\delta\delta'}(\partial^4_{\alpha'\beta'\gamma'\delta'}f)\\
&\quad + xP(\bm{\partial} f)^T\mathcal{R} \bm{\partial}\left[(\partial^3_{\alpha\beta\gamma}f )\mathcal{R}_{\alpha\alpha'}\mathcal{R}_{\beta\beta'} \mathcal{R}_{\gamma\gamma'}(\partial^3_{\alpha'\beta'\gamma'}f)\right]
\end{aligned}
\end{equation}
\begin{equation}
\begin{aligned}
\text{RHS, second term}
&= - P(\partial^3_{\alpha\beta\gamma}D f )\mathcal{R}_{\alpha\alpha'}\mathcal{R}_{\beta\beta'} \mathcal{R}_{\gamma\gamma'}(\partial^3_{\alpha'\beta'\gamma'}f)\\
&\quad - 2xP[ \left(\bm{\partial}\partial^3_{\alpha\beta\gamma}f\right)^T\mathcal{R}\left(\bm{\partial}f\right) ]\mathcal{R}_{\alpha\alpha'}\mathcal{R}_{\beta\beta'} \mathcal{R}_{\gamma\gamma'}(\partial^3_{\alpha'\beta'\gamma'}f)\\
&\quad - 6xP[ \left(\bm{\partial}\partial^2_{\alpha\beta}f\right)^T\mathcal{R}\left(\bm{\partial} \partial_\gamma f\right) ]\mathcal{R}_{\alpha\alpha'}\mathcal{R}_{\beta\beta'} \mathcal{R}_{\gamma\gamma'}(\partial^3_{\alpha'\beta'\gamma'}f)
\end{aligned}
\end{equation}
\begin{equation}
\begin{aligned}
\text{RHS, fourth term}
&=-3x P \Tr [(D\mathcal{H})\mathcal{RHRHR}] - 2x^2 P (\bm{\partial} f )^T\mathcal{R}\bm{\partial} \Tr(\mathcal{H}\mathcal{R})^3\\
&\quad - 6x P \mathcal{R}_{\delta'\delta}(\partial^3_{\alpha \alpha'\delta} f)\mathcal{R}_{\alpha'\beta}(\partial^3_{\beta \beta'\delta'} f)\mathcal{R}_{\beta'\gamma}(\partial^2_{\gamma \gamma'} f)\mathcal{R}_{\gamma'\alpha}
\end{aligned}
\end{equation}
\begin{equation}
\begin{aligned}
\text{RHS, fifth term}
&= 3x P \Tr [(D\mathcal{H})\mathcal{RHRHR}] \\
&\quad+ 6 x^2 P (\left[\bm{\partial} \partial^2_{\alpha\alpha'}f\right]^T\mathcal{R}\left[\bm{\partial}f\right])\mathcal{R}_{\alpha'\beta}(\partial^2_{\beta\beta'}f) \mathcal{R}_{\beta'\gamma}(\partial_{\gamma\gamma'}f)\mathcal{R}_{\gamma'\alpha}\\
&\quad + 6 x^2 P (\left[\bm{\partial} \partial_{\alpha'}f\right]^T\mathcal{R}\left[\partial_{\alpha}\bm{\partial}f\right])\mathcal{R}_{\alpha'\beta}(\partial^2_{\beta\beta'}f) \mathcal{R}_{\beta'\gamma}(\partial_{\gamma\gamma'}f)\mathcal{R}_{\gamma'\alpha}
\end{aligned}
\end{equation}
Summing everything:
\begin{equation}
\begin{aligned}
&\dot{x}\left(\frac{1}{\lambda^3}+\alpha \int \Diff2 \bm{h} \,P \Tr(\mathcal{HR})^3 \right) - \frac{3 x \dot{\lambda}}{\lambda^4}\\
&= \frac{\alpha}{2} \int \Diff2 \bm{h}\,P \Bigl[(\partial^4_{\alpha\beta\gamma\delta} f )\mathcal{R}_{\alpha\alpha'}\mathcal{R}_{\beta\beta'} \mathcal{R}_{\gamma\gamma'}\mathcal{R}_{\delta\delta'}(\partial^4_{\alpha'\beta'\gamma'\delta'}f) 
\\
&\quad- 12 x (\partial^3_{\alpha \beta \gamma } f)\mathcal{R}_{\beta \beta'} \mathcal{R}_{\gamma\gamma'}(\partial^3_{\beta' \gamma'\alpha'} f)\mathcal{R}_{\alpha'\delta}(\partial^2_{\delta \delta'} f)\mathcal{R}_{\delta'\alpha}+6 x^2 \Tr (\mathcal{HR})^4\Bigr]\,.
\end{aligned}
\end{equation}
Again, from Eq.~\eqref{eq:FRSB_saddle3_0} I know that
\begin{equation}
\frac{1}{\lambda^3}+\alpha \int \Diff2 \bm{y}P \Tr(\mathcal{HR})^3 = \frac{\alpha}{2x} \int \Diff2 \bm{y} P (\partial^3_{\alpha\beta\gamma} f )\mathcal{R}_{\alpha\alpha'}\mathcal{R}_{\beta\beta'} \mathcal{R}_{\gamma\gamma'}(\partial^3_{\alpha'\beta'\gamma'}f)
\end{equation}
Substituting this result I get Eq.~\eqref{eq:FRSB_saddle4}.

\section{RS stability}
\label{app:stability}

I report here the main formulas to obtain the results in Sec.~\ref{sec:Stability}.

\subsection{sCSP}
\label{app:stabilityLSO}
For the sCSP, I can write
\begin{equation}
\begin{aligned}
f_\text{s}(q_M,\bm{h}) &= \log \gamma_{1-q_M} \star v_{\text{s}}(\bm{h})\\
&=\log\left[L\!\left(-\frac{h}{\sqrt{1-q_M}},-\frac{\bar{h}}{\sqrt{1-q_M}};\rho\right) \right] \,,
\end{aligned}
\label{eq:app_f_Gardner}
\end{equation}
where the function $L$ is defined as
\begin{equation}
L(h,k;\rho) = \int_h^{+\infty}\int_k^{+\infty} \frac{\diff x\diff y}{2\pi \sqrt{1-\rho^2}} e^{-\frac{x^2 + y^2 -2\rho xy}{2\left(1-\rho^2\right)}}\,.
\label{eq:app_L}
\end{equation}
This integral is well known in last century's mathematical literature. Unfortunately, only special values are known analytically, as (see~\cite{AbramowitzStegun}, Eq.~(26.3.19))
\begin{align}
L(0,0;\rho) &= \frac{1}{4} + \frac{\arcsin \rho}{2\pi}\,,\label{eq:L00}\\
L(h,k;0) &= \frac{1}{2}\erfc\!\left(\frac{h}{\sqrt{2}}\right)\frac{1}{2}\erfc\!\left(\frac{k}{\sqrt{2}}\right)\,.
\end{align}
As it is crucial to evaluate it numerically in a fast way, I use the nice result from~\cite{Drezner:1990}:
\begin{equation}
L(h,k;\rho) = \int_0^{\rho} \diff r\, \partial_r L(h,k;r) +\Phi(-h)\Phi(-k)\,,
\end{equation}
where
\begin{equation}
\partial_\rho L(h,k;\rho) =\frac{1}{2\pi \sqrt{1-\rho^2}} e^{-\frac{h^2 + k^2 -2\rho hk}{2\left(1-\rho^2\right)}}
\end{equation}
and $\Phi$ is the CDF of the standard normal distribution. Using Gaussian quadrature formulas based on $N$ points, it is possible to approximate the integral just evaluating $N$ times the integrand, obtaining a considerable speed up in the numerics; this approximation is poor for $|\rho|\approx 1$. Using coefficients taken from~\cite{AbramowitzStegun}, tab. 25.4, p. 916, I can write, for $N$ even,
\begin{equation}
L(h, k; \rho) \approx  \sum_{i=1}^{N/2}
      \frac{\rho w_i}{2} \left\{\partial_\rho L[h, k; (1 + x_i)\rho /2 ] +
         \partial_\rho L[h, k, (1 - x_i)\rho/2 ]\right\} 
  + \Phi(-h)\Phi(-k)\,.
\end{equation}
For the numerical results in this paper, I use $N=10$.

When $q_M\to 1$, the integral in Eq.~\eqref{eq:app_f_Gardner} can be evaluated by asymptotic analysis. The corresponding leading-order value is
\begin{multline}
f_\text{s}(q_M,\bm{h})\sim-\frac{1}{2(1-q_M)}\bigl[ \theta(h) \theta(\rho h - \bar{h}) h^2
+ \theta(\rho \bar{h} - h) \theta(\bar{h})  \bar{h}^2 \\
+ \theta(h - \rho \bar{h}) \theta(\bar{h}- \rho h) \bm{h}^T \mathcal{R}^{-1}\bm{h} \bigr]\,,
\end{multline}
so that
\begin{multline}
\left[\bm{\partial}f_\text{s}(q_M,\bm{h})\right]^T\mathcal{R}\left[\bm{\partial}f_\text{s}(q_M,\bm{h})\right]\sim\frac{1}{(1-q_M)^2}\bigl[ \theta(h) \theta(\rho h - \bar{h}) h^2
+ \theta(\rho \bar{h} - h) \theta(\bar{h})  \bar{h}^2 \\
+ \theta(h - \rho \bar{h}) \theta(\bar{h}- \rho h) \bm{h}^T \mathcal{R}^{-1}\bm{h} \bigr]
\end{multline}
and
\begin{multline}
\Tr\!\left[\mathcal{H}_\text{s}^{\text{RS}}(q_M,\bm{h})\mathcal{R}\mathcal{H}_\text{s}^{\text{RS}}(q_M,\bm{h})\mathcal{R}\right]\sim\frac{1}{(1-q_M)^2}\bigl[ \theta(h) \theta(\rho h - \bar{h})
+ \theta(\rho \bar{h} - h) \theta(\bar{h}) \\
+ 2\theta(h - \rho \bar{h}) \theta(\bar{h}- \rho h) \bigr]\,.
\end{multline}
Integrating over $\bm{h}$ with the Gaussian measure, I find
\begin{equation}
\begin{aligned}
&(1-q_M)^2\int \frac{\Diff2 \bm{h}}{2\pi q_{M} \sqrt{\det \mathcal R}} e^{-\frac{1}{2q_M} \bm{h}^T \mathcal{R}^{-1}\bm{h}}\left[\bm{\partial}f_\text{s}(q_M,\bm{h})\right]^T\mathcal{R}\left[\bm{\partial}f_\text{s}(q_M,\bm{h})\right]\\
&\quad\sim \frac{2}{2\pi\sqrt{1-\rho^2}} \int_0^{+\infty} \diff h \,h^2 \int_{-\infty}^{\rho h} \diff \bar{h} \, e^{-\frac{1}{2} \bm{h}^T \mathcal{R}^{-1}\bm{h}} \\
&\hphantom{\quad\sim{}}+  \int \frac{\Diff2\bm{h}}{2\pi\sqrt{1-\rho^2}}\theta(h - \rho \bar{h}) \theta(\bar{h}- \rho h) \bm{h}^T \mathcal{R}^{-1}\bm{h} \,e^{-\frac{1}{2} \bm{h}^T \mathcal{R}^{-1}\bm{h}}\\
&\quad \sim \frac{1}{2}+ \frac{\sqrt{1-\rho^2}}{2\pi}\iint_0^{+\infty} \!\Diff2\bm{z} \,\bm{z}^T \mathcal{R}\bm{z}\, e^{-\frac{1}{2} \bm{z}^T \mathcal{R}\bm{z}}\\
&\quad \sim \frac{1}{2} + \frac{2}{\pi}\arctan\!\left(\sqrt{\frac{1-\rho}{1+\rho}}\right)\,
\end{aligned}
\end{equation}
and
\begin{equation}
\begin{aligned}
&(1-q_M)^2\int \frac{\Diff2 \bm{h}}{2\pi q_{M} \sqrt{\det \mathcal R}} e^{-\frac{1}{2q_M} \bm{h}^T \mathcal{R}^{-1}\bm{h}}\Tr\!\left[\mathcal{H}_\text{s}^{\text{RS}}(q_M,\bm{y})\mathcal{R}\mathcal{H}_\text{s}^{\text{RS}}(q_M,\bm{h})\mathcal{R}\right]\\
&\quad\sim \frac{1}{\pi\sqrt{1-\rho^2}}\left[ \int_0^{+\infty}\!\! \diff h  \int_{-\infty}^{\rho h} \diff \bar{h} \, e^{-\frac{1}{2} \bm{h}^T \mathcal{R}^{-1}\bm{h}} +  \int \Diff2\bm{h}\,\theta(h - \rho \bar{h}) \theta(\bar{h}- \rho h) \,e^{-\frac{1}{2} \bm{h}^T \mathcal{R}^{-1}\bm{h}}\right]\\
&\quad\sim \frac{1}{2} +  \frac{1}{\pi\sqrt{1-\rho^2}}\int \Diff2\bm{h}\,\theta(h - \rho \bar{h}) \theta(\bar{h}- \rho h) \,e^{-\frac{1}{2} \bm{h}^T \mathcal{R}^{-1}\bm{h}}\\
&\quad \sim \frac{1}{2} + \frac{2}{\pi}\arctan\!\left(\sqrt{\frac{1-\rho}{1+\rho}}\right)\,,
\end{aligned}
\end{equation}
so these quantities are the same.

\subsection{uCSP}
\label{app:GPR}

In this case,
\begin{equation}
\begin{aligned}
f_\text{u}(q_M,\bm{h}) &= \log \gamma_{1-q_M} \star v_{\text{u}}(\bm{h})\\
&=\log\left[L\!\left(-\frac{h}{\sqrt{1-q_M}},-\frac{\bar{h}}{\sqrt{1-q_M}};\rho\right) + L\!\left(\frac{h}{\sqrt{1-q_M}},\frac{\bar{h}}{\sqrt{1-q_M}};\rho\right) \right] \,,
\end{aligned}
\label{eq:app_f_GPR}
\end{equation}
where the function $L$ is defined in Eq.~\eqref{eq:app_L}. For the asymptotic analysis in the limit $q_M\to 1$, I address the interested reader to~\cite{GPR:2020PRE}. 

To evaluate the integrals appearing in the saddle-point and dAT equations in the special limit $q_M\to 0$, it is useful to apply the change of variables $\bm{h} \to \sqrt{q_M} \bm{h}$ in the outer Gaussian convolution, writing, for example,
\begin{multline}
\int \frac{\Diff2 \bm{h}}{2\pi q_{M} \sqrt{\det \mathcal R}} e^{-\frac{1}{2q_M} \bm{h}^T \mathcal{R}^{-1}\bm{h}}\left[\bm{\partial}f_\text{u}(q_M,\bm{h})\right]^T\mathcal{R}\left[\bm{\partial}f_\text{u}(q_M,\bm{h})\right]\\
=\int \frac{\Diff2 \bm{h}}{2\pi  \sqrt{\det \mathcal R}} e^{-\frac{1}{2} \bm{h}^T \mathcal{R}^{-1}\bm{h}}\left[\frac{1}{\sqrt{q_M}}\bm{\partial}f_\text{u}(q_M,\sqrt{q_M}\bm{h})\right]^T\mathcal{R}\left[\frac{1}{\sqrt{q_M}}\bm{\partial}f_\text{u}(q_M,\sqrt{q_M}\bm{h})\right]\,.
\end{multline}
Now I can expand $f_\text{u}(q_M,\sqrt{q_M}\bm{h})$ for small $q_M$, obtaining
\begin{equation}
\begin{aligned}
f_\text{u}(q_M,\sqrt{q_M}\bm{h})&\approx
\log [2L(0,0;\rho )]+
\frac{q_M \left.\left(h^2 \partial^2_h +2 h\bar{h} \partial_h \partial_{\bar{h}}+\bar{h}^2 \partial^2_{\bar{h}}\right)\!L(h,\bar{h};\rho)\right|_{(0,0;\rho )}}{2 L(0,0;\rho )}\\
&\approx
\log [2L(0,0;\rho )]+
\frac{q_M}{4\pi \sqrt{1-\rho^2} L(0,0;\rho )} \left(-h^2 \rho +2 h \bar{h} -\bar{h}^2 \rho \right)\,,
\end{aligned}
\end{equation}
where I used
\begin{equation}
\begin{aligned}
 \left.\partial_h^2 L(h,\bar{h};\rho)\right|_{0,0;\rho} &= -\int_0^{+\infty} \frac{\diff k}{2\pi \sqrt{1-\rho^2}} \frac{\rho k}{1-\rho^2} e^{-\frac{k^2}{2(1-\rho^2)}} = -\frac{\rho }{2 \pi  \sqrt{1-\rho ^2}}\,,\\
 \left. \partial_h \partial_{\bar{h}}L(h,\bar{h};\rho)\right|_{0,0;\rho} &= \frac{1}{2\pi \sqrt{1-\rho^2}}\,.
\end{aligned}
\end{equation}
Eventually,
\begin{equation}
\frac{1}{q_M}\left[\bm{\partial}f_\text{u}(q_M,\sqrt{q_M}\bm{h})\right]^T\mathcal{R}\left[\bm{\partial}f_\text{u}(q_M,\sqrt{q_M}\bm{h})\right] \\
\approx q_M \frac{ h^2-2 \rho  h \bar{h}+\bar{h}^2}{4 \pi ^2 L(0,0;\rho )^2}
\end{equation}
and Eq.~\eqref{eq:FRSB_RS_GPRdAT} follows.
Moreover,
\begin{equation}
\frac{1}{q_M}\mathcal{H}_\text{u}^\text{RS}(q_M,\sqrt{q_M}\bm{h}) \approx \frac{1}{2\pi \sqrt{1-\rho^2} L(0,0;\rho)}\begin{pmatrix}
-\rho&1\\
1&-\rho
\end{pmatrix}\,,
\end{equation}
meaning that
\begin{align}
\frac{1}{q_M^2}\Tr\{[\mathcal{H}_\text{u}^\text{RS}(q_M,\sqrt{q_M}\bm{h})\mathcal{R}]^2\}&\approx\frac{2 \left(1-\rho ^2\right)^2}{(2\pi)^2 (1-\rho^2) L(0,0;\rho)^2} \,,\\
\frac{1}{q_M^3}\Tr\{[\mathcal{H}_\text{u}^\text{RS}(q_M,\sqrt{q_M}\bm{h})\mathcal{R}]^3\}&\approx 0\,.
\end{align}
I also need
\begin{align}
\left.\partial_h^4 L(h,\bar{h};\rho)\right|_{0,0;\rho} &=-\frac{\rho  \left(2 \rho ^2-3\right)}{2 \pi  \left(1-\rho ^2\right)^{3/2}}\,,\\
\left.\partial^3_{h}\partial_{\bar{h}} L(h,\bar{h};\rho)\right|_{0,0;\rho} &=-\frac{1}{2 \pi  \left(1-\rho ^2\right)^{3/2}}\,,\\
\left.\partial^2_{h}\partial^{2}_{\bar{h}} L(h,\bar{h};\rho)\right|_{0,0;\rho} &=\frac{\rho }{2 \pi  \left(1-\rho ^2\right)^{3/2}}
\end{align}
to expand $f_\text{u}(q_M,\sqrt{q_M}\bm{h})$ to second order in $q_M$ and evaluate the quartic derivatives in Eq.~\eqref{eq:FRSB_RS_GPRslope}. I find
\begin{align}
\frac{1}{q_M^2}\partial_h^4 f_\text{u}(q_M,\sqrt{q_M}\bm{h})&\approx-\frac{2 \pi \rho  \left(2 \rho ^2-3\right) L(0,0;\rho )+3 \rho^2  \sqrt{1-\rho ^2}}{4 \pi ^2 \left(1-\rho ^2\right)^{3/2} L(0,0;\rho )^2}\,,\\
\frac{1}{q_M^2}\partial_h^3\partial_{\bar{h}}^{\vphantom{1}} f_\text{u}(q_M,\sqrt{q_M}\bm{h})&\approx \frac{3 \rho  \sqrt{1-\rho ^2}-2 \pi  L(0,0;\rho )}{4 \pi ^2 \left(1-\rho ^2\right)^{3/2} L(0,0;\rho )^2}\,,\\
\frac{1}{q_M^2}\partial_h^2\partial_{\bar{h}}^{2} f_\text{u}(q_M,\sqrt{q_M}\bm{h})&\approx -\frac{\sqrt{1-\rho ^2} \left(\rho ^2+2\right)-2 \pi  \rho  L(0,0;\rho )}{4 \pi ^2 \left(1-\rho ^2\right)^{3/2} L(0,0;\rho )^2}\,,
\end{align}
so I get, for the sum appearing in Eq.~\eqref{eq:FRSB_RS_GPRslope},
\begin{multline}
[\partial^4_{\alpha\beta\gamma\delta} f_\text{u}(q_M,\bm{h} )]\mathcal{R}_{\alpha\alpha'}\mathcal{R}_{\beta\beta'} \mathcal{R}_{\gamma\gamma'}\mathcal{R}_{\delta\delta'}[\partial^4_{\alpha'\beta'\gamma'\delta'}f_\text{u}(q_M,\bm{h})]\\
\approx\frac{\left(1-\rho ^2\right) \left[6 \pi \rho \sqrt{1-\rho ^2}  L(0,0;\rho )+4 \pi^2 \left(\rho ^2+1\right) L(0,0;\rho )^2-3 \rho ^2+3\right]}{2 \pi ^4 L(0,0;\rho )^4}\,.
\end{multline}

\printbibliography
\end{document}